\documentclass[twocolumn,amsmath,notitlepage,aps,prb,10pt,superscriptaddress,floatfix,letterpaper,reprint
]{revtex4-2}
\usepackage{hyperref}
\usepackage[usenames]{color}
\usepackage{amssymb}
\usepackage{graphicx}
\usepackage{physics}
\usepackage{commath}
\usepackage{todonotes}
\usepackage{eufrak}
\usepackage{bm}

\newcommand{\lati}{\bm{i}}
\newcommand{\latj}{\bm{j}}
\newcommand{\latx}{\bm{x}}
\newcommand{\laty}{\bm{y}}

\newcommand{\lats}{\bm{s}}

\newcommand{\cdag}{\ensuremath{c^\dagger}}

\newcommand{\up}{\uparrow}
\newcommand{\dn}{\downarrow}
\newcommand{\tH}{\bm{H}}

\newcommand{\itee}{\textit{E}}

\begin{document}

\title{Analytical Continuation of  Matrix-Valued Functions: Carath\'{e}odory Formalism}

\author{Jiani Fei}
\affiliation{%
 Department of Physics, University of Michigan, Ann Arbor, Michigan 48109, USA
}%
\author{Chia-Nan Yeh}
\affiliation{%
 Department of Physics, University of Michigan, Ann Arbor, Michigan 48109, USA
}%
\author{Dominika Zgid}
\affiliation{%
 Department of Chemistry, University of Michigan, Ann Arbor, Michigan 48109, USA
}
\affiliation{%
 Department of Physics, University of Michigan, Ann Arbor, Michigan 48109, USA
}%
\author{Emanuel Gull}%
\affiliation{%
 Department of Physics, University of Michigan, Ann Arbor, Michigan 48109, USA
}%

\date{\today}

\begin{abstract}
Finite-temperature quantum field theories are formulated in terms of Green's functions and self-energies on the Matsubara axis.
In multi-orbital systems, these quantities are related to positive semidefinite matrix-valued functions of the Carath\'eodory and Schur class.
Analysis, interpretation and evaluation of derived quantities such as real-frequency response functions requires analytic continuation of the off-diagonal elements to the real axis.
We derive the criteria under which such functions exist for given Matsubara data and present an interpolation algorithm that intrinsically respects their mathematical properties.
For small systems with precise Matsubara data, we find that the continuation exactly recovers all off-diagonal and diagonal elements. In real-materials systems, we show that the precision of the continuation is sufficient for the analytic continuation to commute with the Dyson equation, and we show that the commonly used truncation of off-diagonal self-energy elements leads to considerable approximation artifacts.
Our method paves the way for the systematic evaluation of Matsubara data with equations of many-body theory on the real-frequency axis.
\end{abstract}

\maketitle

\section{Introduction}
The central object of finite-temperature field theory is the Matsubara Green's function $\mathcal{G}_{ij}(i\omega_n)$. Finite temperature simulations ranging from perturbative calculations \cite{Hedin65,Dahlen05,Phillips14} to lattice \cite{Blankenbecler81} and continuous-time \cite{Gull11} quantum Monte Carlo and lattice QCD  \cite{Asakawa01,Tripolt19,Rothkopf20} simulations obtain this quantity. In post-processing, an analytical continuation step to the retarded real axis Green's function $G^R_{ij}(\omega)$ is performed with the purpose of obtaining single-particle excitation spectra, self-energy information, and approximations to susceptibilities that can then directly be related to experiment. 

As small imprecisions in the Matsubara data result in large deviations of the continued quantities, the direct solution of the continuation problem is typically avoided. Instead, methods that aim to fit the Matsubara data with a physically reasonable ({\it i.e.} smooth, positive and normalized) spectral function, such as the maximum entropy analytic continuation \cite{Bryan90,Creffield95,Jarrell96,Beach04,Gunnarsson10,Gunnarsson10B,Bergeron16,Levy17,Gaenko17,Rumetshofer19}, the stochastic analytic continuation (SAC) and variants \cite{Sandvik98,Mishchenko00,Gunnarsson07,Fuchs10,Goulko17,Krivenko19}, the sparse modeling method \cite{Otsuki17,Otsuki20}, or machine learning approaches \cite{Yoon18} are used, from which the real and imaginary parts of the retarded Green's functions, self-energies, and susceptibilities are extracted. These methods work well for single-orbital systems and the diagonal components of Green's functions, especially in the presence of data with statistical uncertainties.

However, Green's functions for all but the simplest quantum systems are matrix-valued objects with both diagonal and off-diagonal entries. While certain quantities such as the total spectral function only depend on the diagonal entries, the analysis of the self-energy, computation of the susceptibilities, or evaluation of equations such as the Dyson equation on the real axis requires knowledge of both diagonal and off-diagonal entries. As the off-diagonal entries may change sign as a function of frequency, standard continuation methods that rely on positivity \cite{Jarrell96} fail.

So far, no reliable and general algorithms for the continuation of matrix-valued Green's functions exist, even though several approaches have been explored. For instance, one may perform a Pad\'e \cite{Vidberg77,Beach00,Gunnarsson10,Ostlin12,Osolin13,Schott16,Han17} continuous fraction interpolation of both the diagonal and off-diagonal terms. However, the continued results typically exhibit continuation artifacts such as negative spectral functions even for the diagonal part.
One may instead transform to a basis that diagonalizes the Green's function or the self-energy for a given Matsubara frequency \cite{Tomczak07,Dang14,Gull14}, neglect the remaining off-diagonal elements, and employ a continuation method for diagonal Green's functions. Unless symmetry dictates that all off-diagonal elements for all frequencies must be zero, this is an uncontrolled approximation and results will depend on the basis chosen. 
Finally, one may generalize the maximum entropy method to matrix-valued functions and off-diagonal terms \cite{Kraberger17,Sim18}, using either a positive-negative or a maximum `quantum' entropy approach. These methods enforce the positive semidefiniteness of the Green's function but eliminate sharp and high-energy features, such as the band structure contained in Green's function data, due to the intrinsic limitations of the fitting procedure.

As we show in this paper, the application of Nevalinna theory~\cite{Fei21} to matrix-valued functions overcomes these limitations, leading to basis-indepedent interpolations that intrinsically respect the analytic structure of matrix-valued Green's functions. Using a generalization of Nevanlinna-Pick interpolation to the class of Carath\'eodory functions, we demonstrate that the analytic continuation of Green's functions, self-energies, and cumulants \cite{Tudor06} leads to matrix-valued functions that are indistinguishable from the exact results for a simple model system.  In the context of a real-materials multi-orbital simulation we then demonstrate that the analytic continuation step commutes with the Dyson equation, such that self-energy, cumulant, and Green's function continuations yield consistent results. Finally, we assess the quality of the frequently used approximation of neglecting off-diagonal self-energy components.

\section{Theory} \label{sec:theory}
The main mathematical objects considered in this paper are matrix-valued Green's functions, self-energies, and cumulants. As we will show below, the analytical properties of these functions ensure that they lie, up to a factor of $i=\sqrt{-1}$,  in the class of Carath\'{e}odory functions \cite{Caratheodory07}.

In the following, we will present the mathematical framework by introducing Carath\'{e}odory functions and their properties, along with the related class of Schur functions \cite{Schur18} and a mapping between the two classes. We will then reformulate the problem of analytical continuation as an interpolation problem in the class of Schur or Carath\'{e}odory functions, in analogy to Ref.~\cite{Fei21}. After detailing the conditions under which a Carath\'{e}odory interpolant through given Matsubara points exists, we will derive an interpolation algorithm. Finally, we will show that Green's functions, self-energies, and cumulants are Carath\'{e}odory functions.

\subsection{Carath\'{e}odory and Schur functions}
Consider an open subset $\mathcal{B}$ of the complex plane, such as the unit disk $\mathcal{D}=\{z:\lvert z\rvert < 1\}$ or the upper half of the complex plane $\mathcal{C}^+=\{z: \Im{z} > 0\}$. A matrix-valued function $F(z):\ \mathcal{B}\rightarrow\mathbb{C}^{m\times m}$, holomorphic on $\mathcal{B}$, belongs to the class of Carath\'{e}odory functions $\mathfrak{C}$ if, for any $z\in\mathcal{B}$, the Hermitian matrix $(F(z)+F^\dagger(z))/2$ is positive semi-definite (PSD) \cite{Caratheodory07, Kamp79, Arov08}.

Analogously, a matrix-valued function $\Psi(z),\ \mathcal{D}\rightarrow \mathbb{C}^{m\times m}$, holomorphic on $\mathcal{D}$,  belongs to the Schur class $\mathcal{S}$ if $\lVert \Psi(z)\rVert \leq 1$ for any $\lvert z\rvert < 1$, where $\lVert \Psi\rVert$ denotes norm of the matrix $\Psi$ \cite{Schur18, Fritzsche12}. We use the spectral norm as the matrix norm in this paper, {\it i.e.} the largest eigenvalue of the constant matrix $\Psi_s:=[\Psi\Psi^\dagger]^{1/2}$, with $(\cdot)^{1/2}$ indicating the Hermitian square root.

While the domain of interest for many-body objects such as the Green's function is the upper half of the complex plane $\mathcal{C}^+$, the traditional mathematical literature mostly considers functions on the open unit disk $\mathcal{D}$. A M\"{o}bius transform $h:\mathcal{C}^+\rightarrow\mathcal{D}, h(z)=\frac{z-i}{z+i}$ maps $\mathcal{C}^+$ to $\mathcal{D}$, and its inverse $h^{-1}(z):\mathcal{D}\rightarrow \mathcal{C}^+$ maps the unit disk back to the upper half of the complex plane.

As shown in detail in \cite{Kamp79, Fritzsche12}, every Carath\'{e}odory function $F(z)\in \mathfrak{C}$ restricted to $\mathcal{D}$ can be mapped to a corresponding Schur function $\Psi(z)\in \mathcal{S}$ with the Cayley transform 
\begin{align}
\Psi(z)=[I-F(z)][I+F(z)]^{-1}\label{transform1},
\end{align}
and its inverse
\begin{align}
F(z)= [I+\Psi(z)]^{-1}[I-\Psi(z)]\label{transform2}.
\end{align}

\begin{figure}[tbh]
\includegraphics[width=\columnwidth]{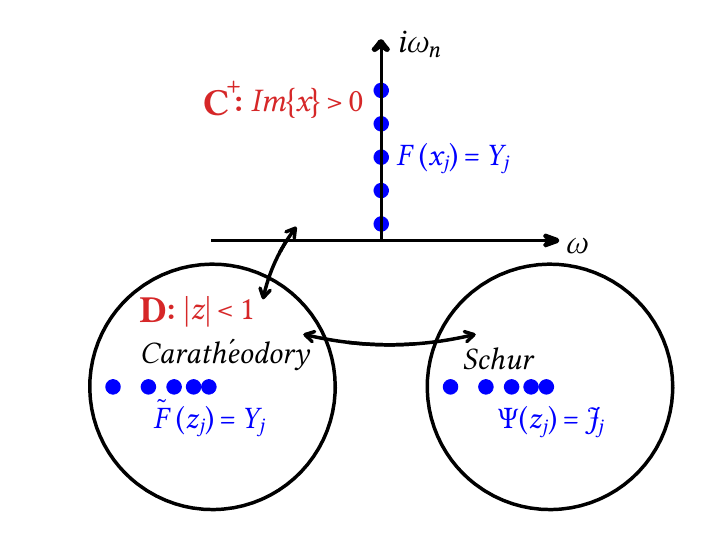}
\caption{Mapping of the input Carath\'{e}odory problem to a Schur interpolation problem. $x_j$ denote Matsubara frequencies, $Y_j$ Matsubara values, $z_j$ transformed Matsubara points, and $J_j$ transformed Matsubara values}\label{fig:sketch}
\end{figure}

As illustrated in Fig.~\ref{fig:sketch}, the construction of a Carath\'{e}odory interpolant on the upper half of the complex plane is therefore equivalent to the construction of a Carath\'{e}odory interpolant on the unit disk, which in turn is equivalent to the construction of the corresponding Schur interpolant on the unit disk.

\subsection{Pick Criterion}
We aim to interpolate matrix-valued Green's functions, self-energies, and cumulants obtained in Matsubara frequencies. Thus, after mapping to the unit disk, the interpolation problem is specified by a set of $n$ Carath\'{e}odory $m\times m$ matrices $Y_j(z_j)$ at $n$ points $z_j \in \mathcal{D}$ (see Fig.~\ref{fig:sketch}). We first consider the conditions under which such an interpolation problem has solutions.

Generalizing the Pick criterion for scalar functions \cite{Pick,Akhiezer,Tannenbaum17} to the matrix-valued case, Refs.~\cite{Kamp79, Chen94} derived an existence criterion for Carath\'{e}odory interpolants directly based on input data. Solutions for the interpolation problem exist if and only if the Pick matrix defined for the Carath\'{e}odory function on the unit disk $\mathcal{D}$,
\begin{align}
P_C=[\frac{Y_k+Y_l^*}{1-z_k^*z_l}]_{(mn)\times(mn)}\label{PickC}
\end{align}
or alternatively, the Pick matrix defined for the transformed Schur function on $\mathcal{D}$,
\begin{align}
P_S=[\frac{I-J_k^*J_l}{1-z_k^*z_l}]_{(mn)\times(mn)}\label{PickS}
\end{align}
is positive semi-definite; and a unique solution only exists if it is singular.

We note that this criterion is very restrictive in practice. Numerical noise in Monte Carlo typically leads to negative eigenvalues such that a Carath\'{e}odory (or Nevanlinna \cite{Fei21}) solution of the interpolation problem does not exist. Round-off and convergence issues in semi-analytical calculations such as GW simulations of real materials (see Sec.~\ref{sec:results}) lead to eigenvalues that are small and negative, such that  the evaluation just above the real axis still leads to PSD spectral functions, while analytical solutions such as those for the Hubbard dimer presented below always satisfy the Pick criterion.

\subsection{Interpolation of Schur functions}
Matrix-valued functions in the Carath\'{e}odory class have continued fraction expansions \cite{Kamp79}. An algorithm to obtain such interpolation is given by matrix extensions of the classical Schur algorithm \cite{Kamp79, Chen94, Cuneyt13}, which we use as interpolation method in this paper.

Given input data $F(x_j)=Y_j$ ($j=0,1,\dots,n-1$; $x_j\in\mathcal{C}^+$; $Y_j\in\mathbb{C}^{m\times m}$) for a Carath\'{e}odory matrix-valued function $F(z)$, we M\"{o}bius transform the domain and conformally map the function value according to Eq. \ref{transform1} (see Fig.~\ref{fig:sketch}), and reformulate the problem as a problem of finding a matrix-valued function $\Psi(z)$ in the Schur class, such that
\begin{align}
\Psi(z_j)&=J_j=\Psi(\frac{x_j-i}{x_j+i})=[I-Y_j][I+Y_j]^{-1}, \nonumber \\
j&=0,1,\dots,n-1.\label{interpolation}
\end{align}

We will proceed as follows. Given a interplation problem $\Psi=\Psi_0$ with $n$ nodes $\Psi(z_j)=J_j$, $j=0,1,\dots,n-1$, we will find a function that interpolates the first node $z_0$ and express the remaining problem as an interpolation problem $\Psi_1$ through $n-1$ nodes. This interpolation problem will then be expressed as a function that interpolates the second node $z_1$ and a remaining problem $\Psi_2$ through $n-2$ nodes. The procedure will be repeated until all nodes are interpolated and only a free Schur function $\Psi_{n}$ remains.

We first show how to reduce the Schur function $\Psi_i(z)\in \mathcal{S}$ to $\Psi_{i+1}(z)\in \mathcal{S}$, while releasing the node constraint of $\Psi_i(z)$ at $z_i$. The reduction step is based on the theories of J-contractive transformations \cite{Potapov55} and as follows \cite{Kamp79}. Assuming $\Psi_i(z_i)=W_i$, define the matrix-valued function $L_i(z)$ by
\begin{align}
y_i L_i(z) =& [I-W_iW_i^\dagger]^{-1/2}[\Psi_i(z)-W_i]\cdot \nonumber \\
&[I-W_i^\dagger \Psi_i(z)]^{-1}[I-W_i^\dagger W_i]^{1/2}\label{iter1}
\end{align}
where $y_i=\lvert z_i\rvert (z_i-z)/(z_i(1-z_i^* z))$. $L_i(z)\in \mathcal{S}$ by the Schwarz lemma \cite{Potapov55}. Define $\Psi_{i+1}(z)$ as
\begin{align}
\Psi_{i+1}(z)=&[I-K_iK_i^\dagger]^{-1/2}[L_i(z)-K_i]\cdot \nonumber \\
&[I-K_i^\dagger L_i(z)]^{-1}[I-K_i^\dagger K_i]^{1/2}\label{iter2}
\end{align}
where $K_i$ is an arbitrary matrix such that $\lVert K_i\rVert < 1$. Since for any $z_k,z_l\in \mathcal{D}$, the matrices $\left[I-L(z_k)^\dagger L(z_l)\right]$ and $\left[I-\Psi_{i+1}(z_k)^\dagger \Psi_{i+1}(z_l)\right]$ are equivalent under similarity transformation, $\Psi_{i+1}(z)\in \mathcal{S}$ as well. 

Notice that in Eq. \ref{iter1}, left- and right-hand-side are both zero at node $z_i$ so that the value of $L_i(z_i)$, and in turn of $\Psi_{i+1}(z_i)$, is free; and that for all remaining nodes $z_j, j>i$, $\Psi_{i+1}(z_j)$ is completely determined by $L_i(z_j)$, in turn by $\Psi_i(z_j)$, up to a freedom in the choice of $K_i$. We therefore have a new interpolation problem $\Psi_{i+1}\in \mathcal{S}$ with one less node constraint. Iterating the complete algorithm backwards, an arbitrary Schur function $\Psi_n(z)\in \mathcal{S}$ will yield $\Psi_0(z)\in \mathcal{S}$ that hits all interpolation nodes. 

$K_i$, $i=0,1,\dots,n-1$ and $\Psi_{n}$ are free parameters that can be used to enforce additional conditions, such as smoothness \cite{Fei21}, and cover all possible interpolants in the class of Schur functions. For convenience, we choose $\Psi_n$ to be the identity matrix and $K_i$ to be zero, for all $i$.

Denoting $\Psi_i(z_j)=W_j^i$, $i, j=0,1,\dots,{n-1}$ and $W_i^i=W_i$, consistent with the above notation, the first stage of the algorithm consists of computing all $W_i$ and storing them. By Eq. (\ref{iter1}) and (\ref{iter2}), we have
\begin{align}
W_j^{i+1}=&\frac{z_i(1-z_i^* z_j)}{\lvert z_i\rvert (z_i-z_j)}[1-W_iW_i^\dagger]^{-1/2}[W_j^i-W_i]\times \nonumber \\
&[1-W_i^\dagger W_j^i]^{-1}[1-W_i^\dagger W_i]^{1/2}\quad j\geq i+1
\end{align}
Iterating through $i=0,\dots,n-2$, we obtain all $W_i$.

The second stage of the algorithm consists of solving $\Psi_{n}(z)\rightarrow \Psi_{n-1}(z)\dots\rightarrow\Psi_{0}(z)=\Psi(z)$ where $\Psi(z), z\in\mathcal{D}$ is the desired Schur interpolant. This is done by transcribing Eq. (\ref{iter1}) and (\ref{iter2}) as
\begin{align}
V_i(z)=&\frac{\lvert z_i\rvert (z_i-z)}{z_i(1-z_i^* z)}\\ \nonumber
&\times [I-W_iW_i^\dagger]^{1/2}\Psi_{i+1}(z)[I-W_i^\dagger W_i]^{-1/2}\\
\Psi_i(z)=&[I+V_i(z)W_i^\dagger]^{-1}[V_i(z)+W_i]
\end{align}
with $i$ iterating from $n-1$ to $0$.

Lastly, we map the Schur class fuctions back to the Carath\'{e}odory space by constructing the Carath\'{e}odory interpolant $F(x)$ via
\begin{align}
F(x)= [I+\Psi(\frac{x-i}{x+i})]^{-1}[I-\Psi(\frac{x-i}{x+i})]\label{transform}
\end{align}
The first stage is computed only once, while the second stage depends on $z$ and is repeated by Eq.~\ref{transform} for each $x\in\mathcal{C}^+$ that needs to be evaluated.

We note that, as in most Pad\'{e} codes, our numerical interpolation is highly sensitive to numerical roundoff and is therefore performed in $512$ or $1024$-bit precision.

\begin{figure}[tbh]
\includegraphics[width=\columnwidth]{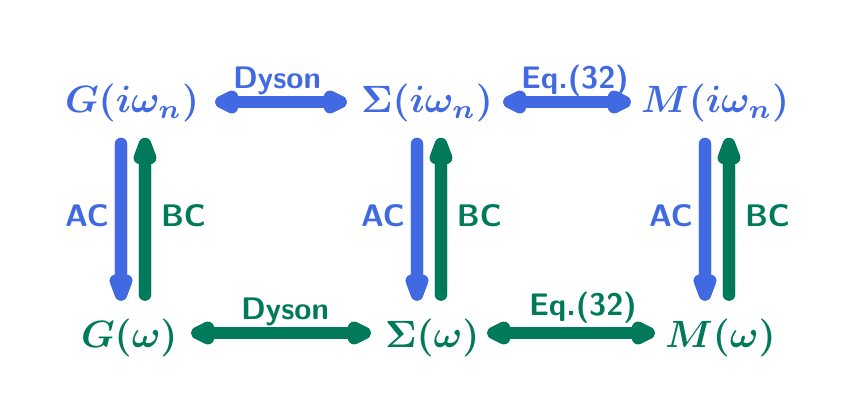}
\caption{Schematic of the transformations between $G(i\omega_n)$, $G(\omega)$, $\Sigma(i\omega_n)$, $\Sigma(\omega)$, $M(i\omega_n)$ and $M(\omega)$. AC stands for analytical continuation. BC stands for back continuation.}\label{fig:schematic}
\end{figure}

\subsection{Carath\'{e}odory Functions in Many-Body Theory}
We now show that the Green's functions, self-energies, and cumulants commonly encountered in many-body theory are Carath\'{e}odory functions (up to factors of $i$) and can therefore be interpolated with the interpolation method described above.
\subsubsection{$iG(z)$ is a Carath\'{e}odory function}
We assume a Hamiltonian system with eigenvectors $\lvert m \rangle$ and eigen energies $\textit{E}_m$. $c_{\lati} ^{\dagger}$ ($c_{\lati}$) is the creation (annihilation) operator for the single-particle orbital $\lati$, $Z=\sum_{m}e^{-\beta\textit{E}_m}$ the partition function and $\beta$ the inverse temperature. 
In the Lehmann representation, the matrix elements of the fermionic Green's function $G(z)$ in the upper half complex plane $\mathcal{C}^+$ are
\begin{align}
	G_{\lati\latj}(z) &= \frac{1}{Z}\sum_{m, n} \frac{\langle n \lvert c_{\lati}\rvert m \rangle \langle m\lvert \cdag_{\latj}\rvert n \rangle}{z + \textit{E}_n - \textit{E}_m}(e^{-\beta \textit{E}_n}+e^{-\beta \textit{E}_m}).\label{eqn: Lehmann} 
\end{align}

We aim to prove for $z\in \mathcal{C}^+$ that $iG(z)+(iG(z))^\dagger$ is a PSD matrix, i.e., $iG(z)$ is a Carath\'{e}odory function on $\mathcal{C}^+$. This follows directly from the definition of a PSD matrix. For any complex vector $\rvert x \rangle$,
\begin{align}
	\langle x \lvert iG(z)&+(iG(z))^\dagger \rvert x \rangle \\
	= \frac{1}{Z}\sum_{mn\lati\latj}&\left[\frac{i(e^{-\beta \itee_m}+e^{-\beta \itee_n})}{z+\itee_n-\itee_m}-\frac{i(e^{-\beta \itee_m}+e^{-\beta \itee_n})}{z^*+\itee_n-\itee_m}\right]\nonumber \\ 
	&\langle n \lvert c_{\lati}x_{\lati}^*\rvert m \rangle \langle m\lvert \cdag_{\latj}x_{\latj}\rvert n \rangle\\
	=\frac{1}{Z}\sum_{mn}&\left[\frac{i(e^{-\beta \itee_m}+e^{-\beta \itee_n})}{z+\itee_n-\itee_m}-\frac{i(e^{-\beta \itee_m}+e^{-\beta \itee_n})}{z^*+\itee_n-\itee_m}\right]\nonumber \\ 
	&\langle n \lvert \sum_{\lati}c_{\lati}x_{\lati}^*\rvert m \rangle^2\\
	=\frac{1}{Z}\sum_{mn}&\frac{2\text{Im}\{z\}\ (e^{-\beta \itee_m}+e^{-\beta \itee_n})}{\text{Im}\{z\}^2+(\text{Re}\{z\}+\itee_n-\itee_m)^2}\nonumber \\ 
	&\langle n \lvert \sum_{\lati}c_{\lati}x_{\lati}^*\rvert m \rangle^2 \geq 0
\end{align}
$iG(z)$ is therefore a Carath\'{e}odory function on $\mathcal{C}^+$.

\subsubsection{$i\Sigma(z)$ is a Carath\'{e}odory function}
In order to show that self-energies are Carath\'{e}odory functions, we make use of the Lehmann representation of the self-energy $\Sigma(z)$ proposed in Ref.~\cite{Karsten14, Christian15}.

The Lehmann representation of $\Sigma(z)$ is constructed with respect to a general fermionic Hamiltonian
\begin{align}
H(t)=&\sum_{\lati\latj}[T_{\lati\latj}(t)-\mu\delta_{\lati\latj}]c_{\lati}^{\dagger}(t)c_{\latj}(t)\nonumber \\
&+\frac{1}{2}\sum_{\lati\latj\lati '\latj '}U_{\lati\lati '\latj\latj '}(t)c_{\latj}(t)c_{\lati'}^{\dagger}(t)c_{\latj'}(t)c_{\lati}(t),\label{general Ham}
\end{align}
where $c_{\lati}(t)=U^\dagger(t,0)c_{\lati}U(t,0)$, $U(t,t')=\mathcal{T}_C\exp(-i\int_t^{t'}H(t_1)dt_1)$ is the system's time-evolution operator and $\mathcal{T}_C$ is the time-ordering operator along the Keldysh-Matsubara contour ($t=0\rightarrow\infty\rightarrow 0 \rightarrow -i\beta$; note that we restrict ourselves here to time-translation invariant systems).

For the self-energy, additional bath degrees of freedom (denoted $\lats$) are added to the physical orbitals (denoted $\lati$, $\latj$; all orbitals denoted $\latx$, $\laty$) in order to emulate the retardation effect (see Ref. \cite{Christian15} for details),
\begin{align}
H_{\text{eff}}(t)=\sum_{\latx\laty} h_{\latx\laty}(t)c_{\lati}^{\dagger}c_{\latj}
\end{align}
$H_{\text{eff}}$ is determined uniquely by making the $\lats\times\lats$ virtual sector diagonal.

The explicit construction of the Lehmann representation of $\Sigma(z)$ for $H(t)$ is as follows \cite{Christian15}, 
\begin{align}
\Sigma_{\lati\latj}(t,t') &= \delta_C(t,t')\Sigma_{\lati\latj}^{\text{HF}}(t)+\Sigma_{\lati\latj}^{C}(t,t') \\
\Sigma_{\lati\latj}^{\text{HF}}(t)&\equiv 2\sum_{\lati'\latj'}U_{\lati\lati '\latj\latj '}(t)\langle\mathcal{T}_C\ c_{\lati'}^{\dagger}(t)c_{\latj'}(t)\rangle _{H_{\text{eff}}}\\
\Sigma_{\lati\latj}^{C}(t,t')&\equiv\sum_{\lats}h_{\lati\lats}(t)g(h_{\lats\lats};t,t')h_{\latj\lats}^*(t').
\end{align}
where the correlated $\Sigma^C$ term is the self-energy of the effective model; $g(\epsilon;t,t') = i[1/(e^{\beta\epsilon}+1)-\Theta_C(t,t')]e^{i\epsilon(t-t')}$ [$\Theta_C(t,t')=1$ for $t\geqslant_C t'$, $\Theta_C(t,t')=0$ otherwise] is the non-interacting Green's function of an isolated one-particle mode ($h_{\text{mode}}=\epsilon c^\dagger c$) with excitation energy $\epsilon$.

We aim to prove that for $z\in \mathcal{C}^+$, $i\Sigma(z)+(i\Sigma(z))^\dagger$ is a PSD matrix, i.e., $i\Sigma(z)$ is a Carath\'{e}odory function on $\mathcal{C}^+$. Fourier transforming $\Sigma^C$ from time to frequency yields
\begin{align}
\Sigma^C_{\lati\latj}(i\omega_n)&=\int_{0}^{\beta}-i\Sigma^C_{\lati\latj}(-i\tau,0)e^{i\omega_n\tau}d\tau\\
&=\int_{0}^{\beta}\sum_{\lats}h_{\lati\lats}(0)h^*_{\latj\lats}(0)\frac{-e^{(\beta + \tau) h_{\lats\lats}}}{e^{\beta h_{\lats\lats}}+1}e^{i\omega_n\tau}d\tau\\
&=\sum_{\lats} \frac{h_{\lati\lats}(0)h^*_{\latj\lats}(0)(e^{i\omega_n\beta}-e^{h_{\lats\lats}\beta})}{(1+e^{h_{\lats\lats}\beta})(h_{\lats\lats}-i\omega_n)}\\
&=\sum_{\lats}\frac{h_{\lati\lats}(0)h^*_{\latj\lats}(0)}{i\omega_n-h_{\lats\lats}}\\
\Sigma^C_{\lati\latj}(z)&=\sum_{\lats}\frac{h_{\lati\lats}(0)h^*_{\latj\lats}(0)}{z-h_{\lats\lats}}
\end{align}
where $\omega_n=\frac{(2n+1)\pi}{\beta}$ are the fermionic Matsubara frequencies so that $e^{i\omega_n\beta}=-1$. Observe that $\Sigma^{\text{HF}}(z)$ is Hermitian and $z$-independent and that self-energy has the property $\Sigma(x+yi)=(\Sigma(x-yi))^\dagger$ for $x, y>0$,  for any complex vector $\rvert x \rangle$,
\begin{align}
&\langle x \lvert i\Sigma(z)+(i\Sigma(z))^\dagger \rvert x \rangle \\
=&\langle x \lvert i\Sigma^C(x+yi)-i\Sigma^C(x-yi) \rvert x \rangle \\
=&\sum_{\lats}\frac{2y\sum_{\lati\latj}x_{\lati}x^*_{\latj}h_{\lati\lats}(0)h^*_{\latj\lats}(0)}{(h_{\lats\lats}-x)^2+y^2} \\
=&\sum_{\lats} \frac{2y(\sum_{\lati}x_{\lati}h_{\lati\lats}(0))^2}{(h_{\lats\lats}-x)^2+y^2}\geqslant 0
\end{align}
$i\Sigma(z)$ is therefore a Carath\'{e}odory function on $\mathcal{C}^+$.

\subsubsection{$iM(z)$ is a Carath\'{e}odory function }
The Cumulant $M$ is a derived object that has the properties of a Green’s function but lacks the `Fock' contribution of the one-body Hamiltonian \cite{Tudor06,Shiro12}. It has the definition of
\begin{align}
M^{-1}(z) &= G^{-1}(z)+F\\
&= (z+\mu)I - \Sigma(z)
\end{align}
where $F$ is the Fock matrix. It is related to the Green's function as
\begin{align}
&-iG^{-1}(z)+i(G^{-1}(z))^{\dagger} \\
=&-iM^{-1}(z)+iF+i(M^{-1}(z))^{\dagger}-iF^{\dagger}\\
=&-iM^{-1}(z)+i(M^{-1}(z))^{\dagger}
\end{align}
Since $F$ is a constant Hermitian matrix, and the inverse of a Carath\'{e}odory function (if it exists) is a Carath\'{e}odory function \cite{Johnson72}, $iM(z)$ is a matrix-valued Carath\'{e}odory function on $\mathcal{C}^+$.

\begin{figure}[tb]
\includegraphics[width=\columnwidth]{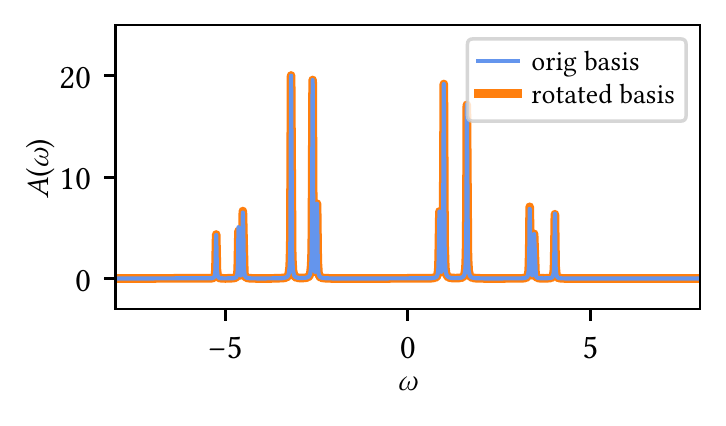} \vspace{-1cm}
\caption{Total spectral function of the Hubbard dimer, obtained in the site basis and in a randomly rotated basis, illustrating the basis independence of the continuation procedure.}\label{fig:rotate}
\end{figure}

\begin{figure*}[tbh]
\includegraphics[width=0.95\textwidth]{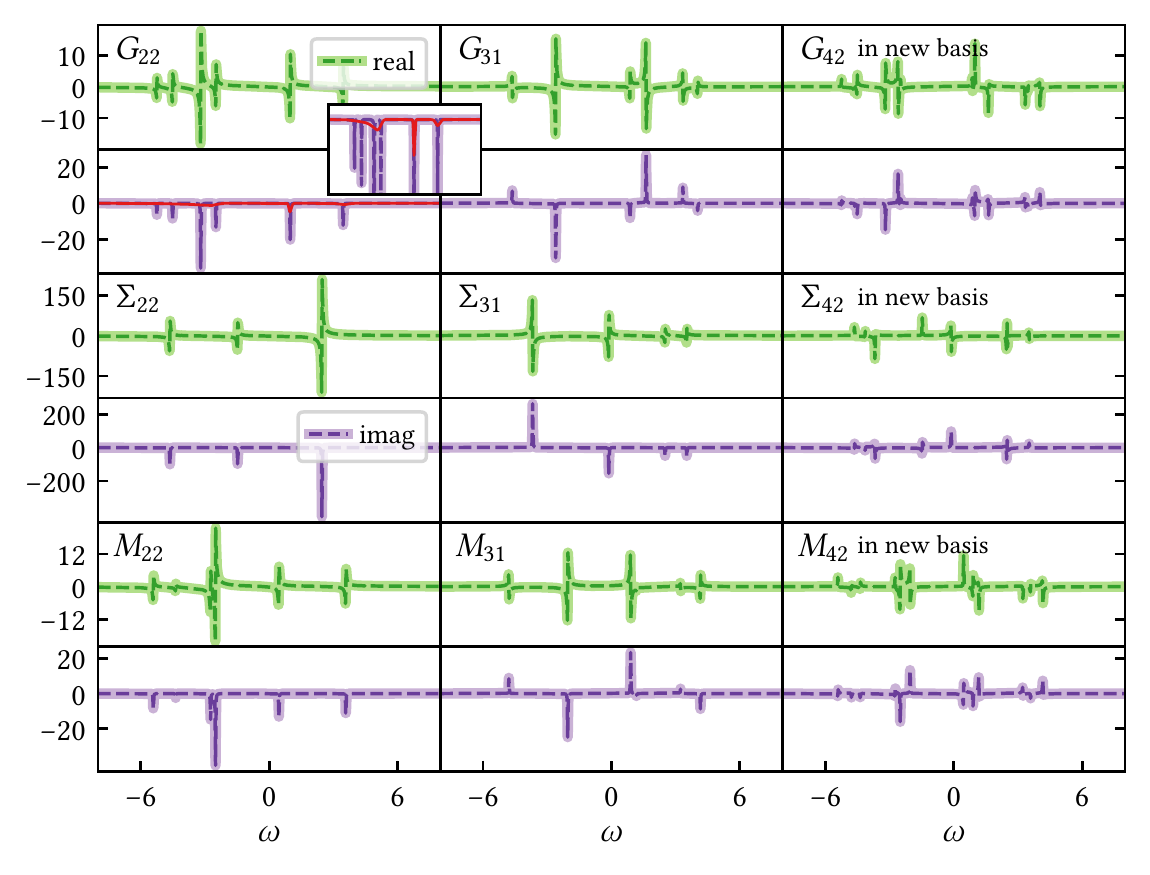}
\caption{\label{fig:dimer} Diagonal and off-diagonal entries of Green's functions, self-energies, and cumulants of a Hubbard dimer (see text for parameters). Dashed lines: exact data evaluated at $\omega+i\eta$ for $\eta=0.01$. Solid lines: analytically continued data evaluated at $\eta=0.01$. Inset and red line: Maximum entropy continuation of $G_{22}$. Top Row: Green's function. Middle Row: Self-energy. Bottom Row: Cumulant. First and second column: diagonal and off-diagonal elements obtained in site-basis. Third column: Off-diagonal entry of data rotated by a random unitary matrix.}
\end{figure*}

\section{Results} \label{sec:results}
\subsection{Hubbard Dimer}
We now show results for analytical continuations. As a first example we choose the Hubbard dimer, which consists of two fermionic spin $1/2$ sites (`0' and `1') which may be occupied by a particle with either spin $\uparrow$ or $\downarrow$. The Fock space of the system contains sixteen configurations, and exact solutions for Green's functions or self-energies can readily be obtained.

The Hamiltonian of the system is 
\begin{align}
H = H_0+H_V+H_{\tH}+H_{SB},
\end{align}
where $H_0 =-\sum_{\sigma}t(\cdag_{0\sigma}c_{1\sigma} + h.c.) - \sum_{i,\sigma} \mu n_{i\sigma}$ describes the usual quadratic spin-diagonal hopping and chemical potential terms and $H_V = \sum_{i} U D_i - \sum_{i,\sigma} \frac{U}{2} n_{i\sigma}$ a Hubbard-type \cite{Qin21} on-site interaction, with $D_i$ denoting the double occupancy on site $i$.
In addition to these terms, we  add a magnetic field and a symmetry-breaking term $H_{\tH} = \sum_{i} \tH(n_{i\up}-n_{i\dn})$ and 
$H_{SB}  = U_a ( D_0 - D_1) +\mu_a (n_{0\up} + n_{0\dn} - n_{1\up} - n_{1\dn}) + \tH_a (n_{0\up} - n_{0\dn} - n_{1\up} + n_{1\dn})$, with the aim of breaking degeneracies and thereby producing a spectral function with additional features. For concreteness, we choose $\beta = 10$, $t = 1$, $U = 5$, $\mu = 0.7$, $\tH = 0.3$, $U_a= 0.5$, $\mu_a = 0.2$, $\tH_a = 0.03$ and the summations are over $i=0, 1$ and $\sigma=\uparrow, \downarrow$. All Green's function results are evaluated on the Matsubara axis and stored in double precision as an input to the interpolation algorithm.

Fig.~\ref{fig:rotate} shows the spectral function of the system. Fig.~\ref{fig:dimer} shows the real (green) and imaginary (purple) parts of the Green's function (top row), self-energy (middle row), and cumulant (bottom row) as a function of frequency. Shown are a sample diagonal (22) element (left column), a sample offdiagonal (31) element (middle column), and an off-diagonal element that has been  obtained after rotating the fermion operators by a random unitary rotation in orbital space (right column).

All dashed lines were obtained exactly by diagonalizing the system in Fock space, obtaining all eigenvalues and eigenvectors, and subsequently computing the real-frequency Green's function via the Lehmann representation. Once the Green's function was known, the process was repeated for the non-interacting system and self-energies and cumulants were obtained by inverting the Dyson equation in real frequencies.

All colored (green, purple) lines were obtained using the algorithm described in Sec.~\ref{sec:theory}. For the Hubbard dimer the results both for the imaginary and the real parts match the exact results precisely, showing the success of the Carath\'{e}odory continuation method. Data was evaluated on the Matsubara axis on a non-equidistant intermediate representation \cite{Shinaoka17,Li20} grid with $36$ positive Matsubara frequency points, interpolated, and evaluated just above the real frequency axis at $\omega+i\eta$ for $\eta=0.01$.

The two top left panels illustrate existing capabilities: diagonal entries of the Green's function are strictly negative, as Green's functions are Nevanlinna \cite{Fei21}. Standard continuation methods such as the Maximum Entropy method \cite{Jarrell96,Levy17} obtain a broadened version of these spectra (red line in the inset of Fig.~\ref{fig:dimer}), from which the real parts can be obtained via a Kramers Kronig relation. If only spectral functions are desired, diagonal components of the Green's functions are sufficient.

The top middle panel exhibits the new capability developed in this work: the analytic continuation of off-diagonal matrix elements. Shown is $G_{13}$, which is one of many off-diagonal components. As is evident, the imaginary part contains both positive and negative entries, and such Green's functions can therefore not be obtained using standard Maximum Entropy \cite{Jarrell96}. As with the diagonal part, real and imaginary parts are related by a Kramers Kronig transform.

\begin{figure*}[tbh]
\includegraphics[width=\textwidth]{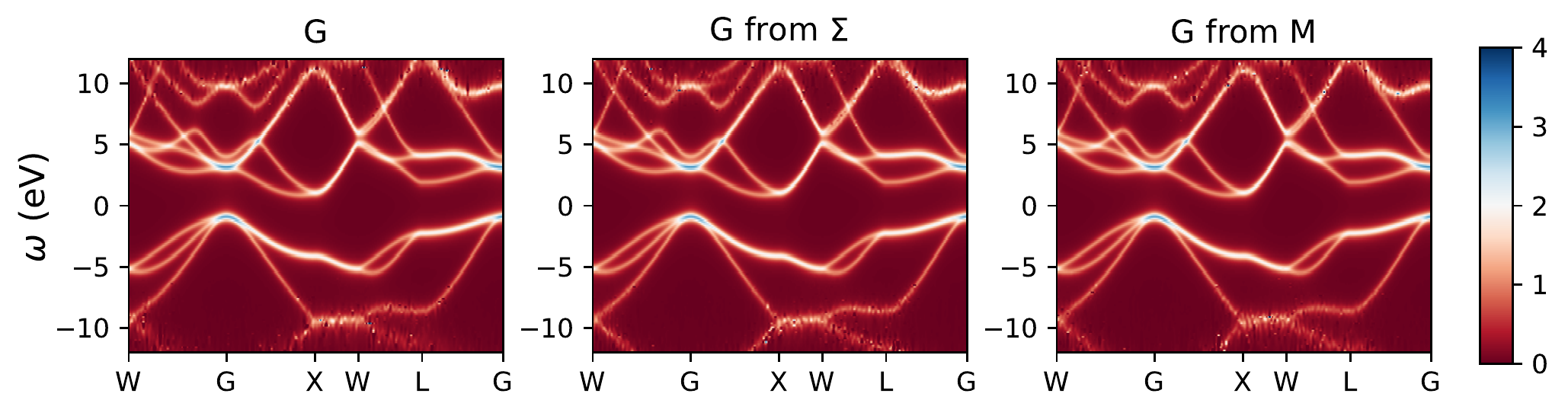}
\caption{Total spectral function of Si evaluated along a high-symmetry path obtained within three approaches, evaluated just above real axis ($\omega+i\eta$ for $\eta=0.01$). Left: continuation of the Green's function $G$. Middle: continuation of the self-energy $\Sigma$ followed by Dyson equation at $\omega+i\eta$. Right: continuation of the cumulant $M$ and application of the Dyson equation at $\omega+i\eta$.}\label{fig:threeG}
\end{figure*}

The top right panel shows a particularly difficult case: an orbital rotation with a random unitary matrix that mixes all Green's function components. While individual elements of the Green's functions are mixed, the rotation preserves the positive definiteness of the Green's function and therefore its Carath\'{e}odory character. The total spectral function, which is basis independent in theory, remains basis independent in practice, giving a first indication that the precision of the continuation and its mathematical properties may be sufficient to perform algebraic operations on the real axis.

To explicitly show that a random orbital rotation leaves the spectral function invariant, Fig.~\ref{fig:rotate} shows the total spectral function of this example, both obtained in the original site basis and in the randomly rotated basis. Peak locations as well as peak heights of the continuations agree after the rotation.

We now turn our attention to the middle row of Fig.~\ref{fig:dimer}. Shown are continuations (real and imaginary parts) of the self-energy. The middle left panel illustrates the present capabilities: similar to Green's functions, diagonal entries of the self-energy are Carath\'{e}odory, and can therefore be continued with established methods \cite{Wang09}. Real parts can then be obtained via a Kramers Kronig transform. However, knowledge of the diagonal parts is not sufficient for obtaining the spectral function, as the inversion in the Dyson equation mixes diagonal and off-diagonal elements. Similar to the continuation of Green's function, no deviation between the exact answer and the continued spectral function is visible.

The middle panel shows off-diagonal self-energies, which exhibit both positive and negative contributions. The right panel contains the self-energies in a randomly rotated basis chosen to maximize off-diagonal contributions. The fact that off-diagonal entries are accessible now allows one to perform arithmetic operations such as the inversion of the Dyson equation on the real axis.

Finally, the bottom three panels of Fig.~\ref{fig:dimer} show the cumulant  \cite{Tudor06} for this case. The cumulant has the same units as the Green's function and can be thought of as the Green's function of an interacting system `without its band structure contribution'. Given the knowledge of the Fock matrix and the matrix-valued cumulant on the real axis, the spectral functions and self-energies can then be obtained via inversion. As in the case of the self-energies and Green's functions, all structure of the system is recovered.

\subsection{Realistic example: Silicon}
Next, we present results for a realistic ab-initio simulation. As an example we show data for crystalline Si, which is a weakly correlated system for which the band structure is well known. The system is solved within the fully self-consistent GW approximation in Gaussian orbitals (\textit{gth-dzvp-molopt-sr} basis \cite{VandeVondele07} and \textit{gth-pbe} pseudopotential \cite{Goedecker96}), using integrals generated by the pySCF package \cite{pySCF20}. Calculations are performed on a $6\times6\times6$ grid \cite{Iskakov20} with $26$ orbitals per unit cell and interpolated using a Wannier interpolation of Matsubara data \cite{Yeh21} on 52 IR positive frequencies  \cite{Shinaoka17,Li20} on a grid with $200$ $k$-points along a high-symmetry path. The total spectral function of the system, obtained by taking the trace of the $26$ elements of the matrix-valued Carath\'{e}odory interpolation of Sec.~\ref{sec:theory}, is shown in the left panel of Fig.~\ref{fig:threeG}. The broadening parameter of $\eta=0.01$ Ha $\sim 3157K$ used for all continuations here is larger than the temperature of $T=0.001$ Ha $\sim 316K$ and correlation effects. Due to the precision of the continuation, individual bands, band gap, and degeneracies at high-symmetry points and directions are clearly visible. (The indirect band gap of $\sim 2$ eV differs from the experimental band gap of 1.1eV due to the self-consistent GW approximation, basis set effects, and finite size effects). High- and low-frequency values near $\pm$ 10 eV show continuation artifacts. We note that continuations for the spectral function can be obtained directly from the diagonal elements of the Green's functions via Nevanlinna continuation \cite{Fei21}, and do not require continuation of the off-diagonal elements.

The middle panel shows the spectral function obtained from a continuation of the self-energy with all off-diagonal elements, followed by the Dyson equation on the real axis (see Fig.~\ref{fig:schematic}). No deviation from the Green's function data is visible, illustrating the precision of the self-energy continuation. As in the left panel, bands, degeneracies, and band gaps are clearly visible. 

Finally, we show the spectral function obtained from the cumulant in the right panel. In realistic systems, cumulants are more convenient to work with than self-energies and Green's functions, as they have the same units as Green's functions, and as (due to the elimination of the Fock term) most of their structure lies at low energies, thereby increasing the precision of arithmetic operations and analytical continuations. No deviation from the Green's function data is visible, illustrating that the numerical solution of the Dyson equation is possible on the real axis.

\begin{figure}[tb]
\includegraphics[width=0.95\columnwidth]{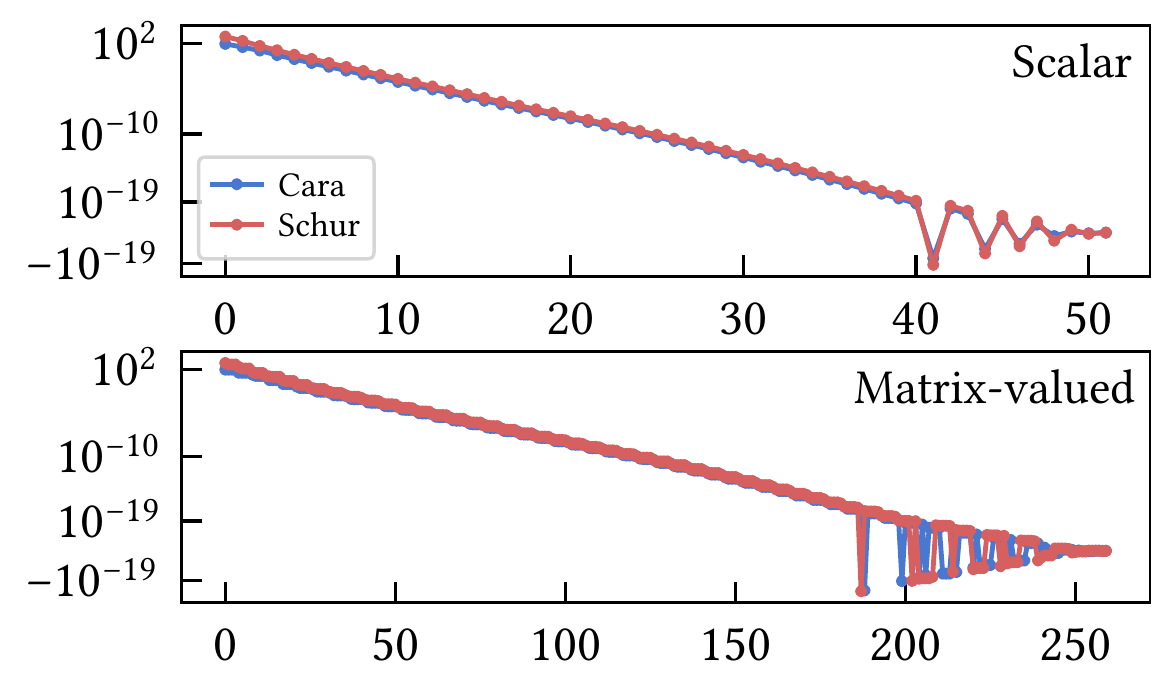}
\caption{\label{fig:Sigma}Eigenvalues of the Pick matrices, Eqs.~\ref{PickC} (blue) and \ref{PickS} (red). Top panel: scalar (Nevanlinna \cite{Fei21}) continuation of the Green's function of Si. Bottom panel: Eigenvalues of the generalized Pick matrix of the matrix-valued problem.}\label{fig:PickMatrix}
\end{figure}

\begin{figure*}[tb]
\includegraphics[width=\textwidth]{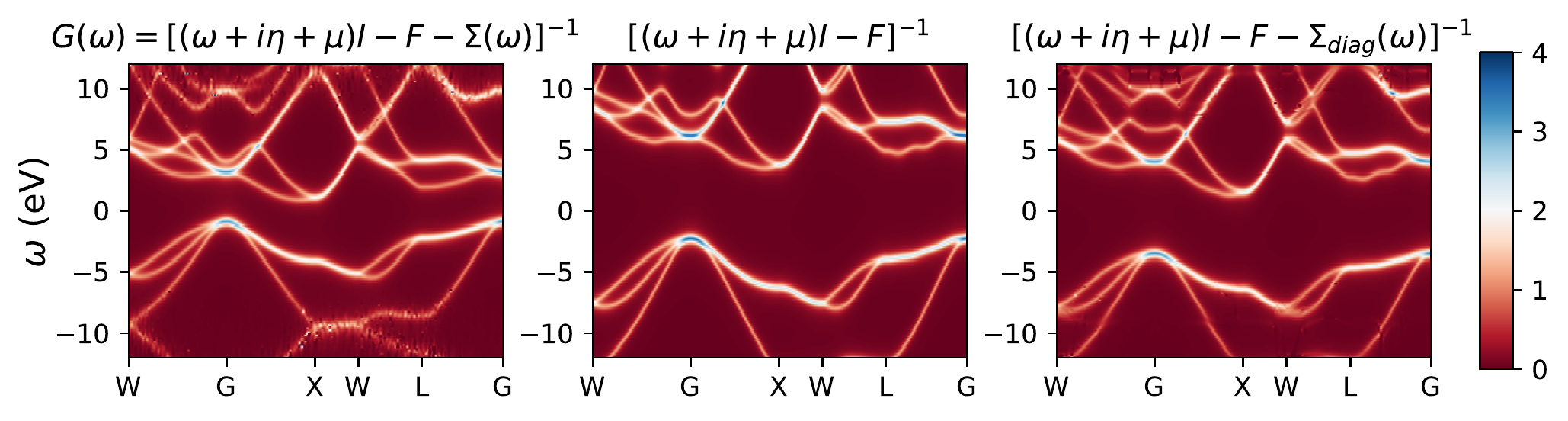}
\caption{\label{fig:threeSigmaG}Effect of the commonly used diagonal approximation of the self-energy on the spectral function. Left panel: Spectral function obtained by continuing the self-energy including all off-diagonal parts;  Middle panel: Spectral function obtained by omitting the dynamical self-energy altogether; Right panel: Spectral function obtained by continuing only the diagonal parts of the dynamical self-energy using the method of Ref.~\cite{Fei21}.}
\end{figure*}

Carath\'{e}odory continuations are sensitive to the precision of the input data. In practice, for large systems, we find that some eigenvalues of the Pick matrix dip below zero due to finite double precision and roundoff errors, and that therefore no Carath\'{e}odory interpolant exists on the unit disk. This is illustrated by Fig.~\ref{fig:PickMatrix}, which  shows the Pick matrix eigenvalues for the Green's function continuations of Fig.~\ref{fig:threeG}. However, even in the case where the Pick criterion is violated, interpolations can still be performed, and the evaluation just above the real axis (in our case shifted by $\eta=0.01$) often leads to positive definite spectral functions. Negative contributions, where they occur, lie very close to the real axis and in frequency regions above the $10$ eV shown here.

\subsection{Diagonal Approximation of the Self-energy}
It is common practice to perform continuations only of the diagonal parts of the self-energy or cumulant, truncating all off-diagonal contributions. In systems where the self-energy can be chosen to be diagonal for all frequencies due to symmetry, such as in rotationally invariant three-orbital systems \cite{Werner09,Georges13,Yin11} and certain model systems, this is exact. However, self-energies are in general off-diagonal, and can only be diagonalized for one Matsubara frequency at a time. Performing a diagonal truncation of the self-energy is an uncontrolled approximation.

To illustrate the effect of the off-diagonal entries of the self-energy on the spectral function, we show in Fig.~\ref{fig:threeSigmaG} results from continuing the self-energy with all entries along with results obtained by only continuing the diagonal part of the dynamical self-energy and results that completely omit the dynamical part of the self-energy. While overall features remain robust, the resulting bands of the diagonal and mean field continuation are shifted and distorted with respect to the full matrix continuation, indicating that a correct treatment of the off-diagonal matrix elements of the self-energy is essential for precise results \cite{Sim18}, even in systems that are only weakly correlated.

\section{Conclusions}
In conclusion, we have shown that matrix-valued Green's functions, self-energies, and cumulants are related to Carath\'eodory and Schur functions. With this analytic structure, we were able to demonstrate the conditions under which a causal positive semi-definite interpolation exists, to construct an interpolation algorithm, and to parametrize all causal interpolations.
An application to a simple benchmark problem showed that diagonal and off-diagonal parts of the self-energies, cumulants, and Green's functions could be recovered to high precision. A demonstration for a real materials multi-orbital problem showed that the analytic continuation commutes with the application of the Dyson equation, and that a diagonal approximation of the self-energy did not produce satisfactory results.

Knowledge of the off-diagonal elements of the real-frequency Green's functions, self-energies, and cumulants is a prerequisite for the evaluation of the equations of many-body theory on the real axis including the real-frequency calculation of susceptibilities and other response functions, and the analysis and inversion of self-energies.

In systems where precise input data is available, our method provides accurate continuations of off-diagonal terms. In system where substantial noise is present, such as in those solved by multi-orbital QMC impurity solvers \cite{Gull11}, a positive definite interpolation does not exist. Causal projections or generalizations of the Maximum entropy fitting procedure \cite{Kraberger17,Sim18} may then be used instead. The derivation of such methods, in addition to the development of methods for the continuation of bosonic quantities, is an interesting problem for future investigation.
 
\begin{acknowledgments}
EG and FJ were supported by the Simons Foundation via the Simons Collaboration on the Many-Electron Problem. DZ and CNY were supported by the U.S. Department of Energy under Award No. DE-SC0019374.
\end{acknowledgments}

\bibliographystyle{apsrev4-2}
\bibliography{refs}

\begin{thebibliography}{67}%
\makeatletter
\providecommand \@ifxundefined [1]{%
 \@ifx{#1\undefined}
}%
\providecommand \@ifnum [1]{%
 \ifnum #1\expandafter \@firstoftwo
 \else \expandafter \@secondoftwo
 \fi
}%
\providecommand \@ifx [1]{%
 \ifx #1\expandafter \@firstoftwo
 \else \expandafter \@secondoftwo
 \fi
}%
\providecommand \natexlab [1]{#1}%
\providecommand \enquote  [1]{``#1''}%
\providecommand \bibnamefont  [1]{#1}%
\providecommand \bibfnamefont [1]{#1}%
\providecommand \citenamefont [1]{#1}%
\providecommand \href@noop [0]{\@secondoftwo}%
\providecommand \href [0]{\begingroup \@sanitize@url \@href}%
\providecommand \@href[1]{\@@startlink{#1}\@@href}%
\providecommand \@@href[1]{\endgroup#1\@@endlink}%
\providecommand \@sanitize@url [0]{\catcode `\\12\catcode `\$12\catcode
  `\&12\catcode `\#12\catcode `\^12\catcode `\_12\catcode `\%12\relax}%
\providecommand \@@startlink[1]{}%
\providecommand \@@endlink[0]{}%
\providecommand \url  [0]{\begingroup\@sanitize@url \@url }%
\providecommand \@url [1]{\endgroup\@href {#1}{\urlprefix }}%
\providecommand \urlprefix  [0]{URL }%
\providecommand \Eprint [0]{\href }%
\providecommand \doibase [0]{https://doi.org/}%
\providecommand \selectlanguage [0]{\@gobble}%
\providecommand \bibinfo  [0]{\@secondoftwo}%
\providecommand \bibfield  [0]{\@secondoftwo}%
\providecommand \translation [1]{[#1]}%
\providecommand \BibitemOpen [0]{}%
\providecommand \bibitemStop [0]{}%
\providecommand \bibitemNoStop [0]{.\EOS\space}%
\providecommand \EOS [0]{\spacefactor3000\relax}%
\providecommand \BibitemShut  [1]{\csname bibitem#1\endcsname}%
\let\auto@bib@innerbib\@empty
\bibitem [{\citenamefont {Hedin}(1965)}]{Hedin65}%
  \BibitemOpen
  \bibfield  {author} {\bibinfo {author} {\bibfnamefont {L.}~\bibnamefont
  {Hedin}},\ }\href {https://doi.org/10.1103/PhysRev.139.A796} {\bibfield
  {journal} {\bibinfo  {journal} {Phys. Rev.}\ }\textbf {\bibinfo {volume}
  {139}},\ \bibinfo {pages} {A796} (\bibinfo {year} {1965})}\BibitemShut
  {NoStop}%
\bibitem [{\citenamefont {Dahlen}\ and\ \citenamefont {van
  Leeuwen}(2005)}]{Dahlen05}%
  \BibitemOpen
  \bibfield  {author} {\bibinfo {author} {\bibfnamefont {N.~E.}\ \bibnamefont
  {Dahlen}}\ and\ \bibinfo {author} {\bibfnamefont {R.}~\bibnamefont {van
  Leeuwen}},\ }\href {https://doi.org/10.1063/1.1884965} {\bibfield  {journal}
  {\bibinfo  {journal} {The Journal of Chemical Physics}\ }\textbf {\bibinfo
  {volume} {122}},\ \bibinfo {pages} {164102} (\bibinfo {year}
  {2005})}\BibitemShut {NoStop}%
\bibitem [{\citenamefont {Phillips}\ and\ \citenamefont
  {Zgid}(2014)}]{Phillips14}%
  \BibitemOpen
  \bibfield  {author} {\bibinfo {author} {\bibfnamefont {J.~J.}\ \bibnamefont
  {Phillips}}\ and\ \bibinfo {author} {\bibfnamefont {D.}~\bibnamefont
  {Zgid}},\ }\href {https://doi.org/10.1063/1.4884951} {\bibfield  {journal}
  {\bibinfo  {journal} {The Journal of Chemical Physics}\ }\textbf {\bibinfo
  {volume} {140}},\ \bibinfo {pages} {241101} (\bibinfo {year}
  {2014})}\BibitemShut {NoStop}%
\bibitem [{\citenamefont {Blankenbecler}\ \emph {et~al.}(1981)\citenamefont
  {Blankenbecler}, \citenamefont {Scalapino},\ and\ \citenamefont
  {Sugar}}]{Blankenbecler81}%
  \BibitemOpen
  \bibfield  {author} {\bibinfo {author} {\bibfnamefont {R.}~\bibnamefont
  {Blankenbecler}}, \bibinfo {author} {\bibfnamefont {D.~J.}\ \bibnamefont
  {Scalapino}},\ and\ \bibinfo {author} {\bibfnamefont {R.~L.}\ \bibnamefont
  {Sugar}},\ }\href {https://doi.org/10.1103/PhysRevD.24.2278} {\bibfield
  {journal} {\bibinfo  {journal} {Phys. Rev. D}\ }\textbf {\bibinfo {volume}
  {24}},\ \bibinfo {pages} {2278} (\bibinfo {year} {1981})}\BibitemShut
  {NoStop}%
\bibitem [{\citenamefont {Gull}\ \emph {et~al.}(2011)\citenamefont {Gull},
  \citenamefont {Millis}, \citenamefont {Lichtenstein}, \citenamefont
  {Rubtsov}, \citenamefont {Troyer},\ and\ \citenamefont {Werner}}]{Gull11}%
  \BibitemOpen
  \bibfield  {author} {\bibinfo {author} {\bibfnamefont {E.}~\bibnamefont
  {Gull}}, \bibinfo {author} {\bibfnamefont {A.~J.}\ \bibnamefont {Millis}},
  \bibinfo {author} {\bibfnamefont {A.~I.}\ \bibnamefont {Lichtenstein}},
  \bibinfo {author} {\bibfnamefont {A.~N.}\ \bibnamefont {Rubtsov}}, \bibinfo
  {author} {\bibfnamefont {M.}~\bibnamefont {Troyer}},\ and\ \bibinfo {author}
  {\bibfnamefont {P.}~\bibnamefont {Werner}},\ }\href
  {https://doi.org/10.1103/RevModPhys.83.349} {\bibfield  {journal} {\bibinfo
  {journal} {Rev. Mod. Phys.}\ }\textbf {\bibinfo {volume} {83}},\ \bibinfo
  {pages} {349} (\bibinfo {year} {2011})}\BibitemShut {NoStop}%
\bibitem [{\citenamefont {Asakawa}\ \emph {et~al.}(2001)\citenamefont
  {Asakawa}, \citenamefont {Nakahara},\ and\ \citenamefont
  {Hatsuda}}]{Asakawa01}%
  \BibitemOpen
  \bibfield  {author} {\bibinfo {author} {\bibfnamefont {M.}~\bibnamefont
  {Asakawa}}, \bibinfo {author} {\bibfnamefont {Y.}~\bibnamefont {Nakahara}},\
  and\ \bibinfo {author} {\bibfnamefont {T.}~\bibnamefont {Hatsuda}},\ }\href
  {https://doi.org/https://doi.org/10.1016/S0146-6410(01)00150-8} {\bibfield
  {journal} {\bibinfo  {journal} {Progress in Particle and Nuclear Physics}\
  }\textbf {\bibinfo {volume} {46}},\ \bibinfo {pages} {459 } (\bibinfo {year}
  {2001})}\BibitemShut {NoStop}%
\bibitem [{\citenamefont {Tripolt}\ \emph {et~al.}(2019)\citenamefont
  {Tripolt}, \citenamefont {Gubler}, \citenamefont {Ulybyshev},\ and\
  \citenamefont {{von Smekal}}}]{Tripolt19}%
  \BibitemOpen
  \bibfield  {author} {\bibinfo {author} {\bibfnamefont {R.-A.}\ \bibnamefont
  {Tripolt}}, \bibinfo {author} {\bibfnamefont {P.}~\bibnamefont {Gubler}},
  \bibinfo {author} {\bibfnamefont {M.}~\bibnamefont {Ulybyshev}},\ and\
  \bibinfo {author} {\bibfnamefont {L.}~\bibnamefont {{von Smekal}}},\ }\href
  {https://doi.org/https://doi.org/10.1016/j.cpc.2018.11.012} {\bibfield
  {journal} {\bibinfo  {journal} {Computer Physics Communications}\ }\textbf
  {\bibinfo {volume} {237}},\ \bibinfo {pages} {129 } (\bibinfo {year}
  {2019})}\BibitemShut {NoStop}%
\bibitem [{\citenamefont {Rothkopf}(2020)}]{Rothkopf20}%
  \BibitemOpen
  \bibfield  {author} {\bibinfo {author} {\bibfnamefont {A.}~\bibnamefont
  {Rothkopf}},\ }\bibfield  {journal} {\bibinfo  {journal} {Data}\ }\textbf
  {\bibinfo {volume} {5}},\ \href {https://doi.org/10.3390/data5030085}
  {10.3390/data5030085} (\bibinfo {year} {2020})\BibitemShut {NoStop}%
\bibitem [{\citenamefont {Bryan}(1990)}]{Bryan90}%
  \BibitemOpen
  \bibfield  {author} {\bibinfo {author} {\bibfnamefont {R.~K.}\ \bibnamefont
  {Bryan}},\ }\href {https://doi.org/10.1007/BF02427376} {\bibfield  {journal}
  {\bibinfo  {journal} {European Biophysics Journal}\ }\textbf {\bibinfo
  {volume} {18}},\ \bibinfo {pages} {165} (\bibinfo {year} {1990})}\BibitemShut
  {NoStop}%
\bibitem [{\citenamefont {Creffield}\ \emph {et~al.}(1995)\citenamefont
  {Creffield}, \citenamefont {Klepfish}, \citenamefont {Pike},\ and\
  \citenamefont {Sarkar}}]{Creffield95}%
  \BibitemOpen
  \bibfield  {author} {\bibinfo {author} {\bibfnamefont {C.~E.}\ \bibnamefont
  {Creffield}}, \bibinfo {author} {\bibfnamefont {E.~G.}\ \bibnamefont
  {Klepfish}}, \bibinfo {author} {\bibfnamefont {E.~R.}\ \bibnamefont {Pike}},\
  and\ \bibinfo {author} {\bibfnamefont {S.}~\bibnamefont {Sarkar}},\ }\href
  {https://doi.org/10.1103/PhysRevLett.75.517} {\bibfield  {journal} {\bibinfo
  {journal} {Phys. Rev. Lett.}\ }\textbf {\bibinfo {volume} {75}},\ \bibinfo
  {pages} {517} (\bibinfo {year} {1995})}\BibitemShut {NoStop}%
\bibitem [{\citenamefont {Jarrell}\ and\ \citenamefont
  {Gubernatis}(1996)}]{Jarrell96}%
  \BibitemOpen
  \bibfield  {author} {\bibinfo {author} {\bibfnamefont {M.}~\bibnamefont
  {Jarrell}}\ and\ \bibinfo {author} {\bibfnamefont {J.}~\bibnamefont
  {Gubernatis}},\ }\href
  {https://doi.org/https://doi.org/10.1016/0370-1573(95)00074-7} {\bibfield
  {journal} {\bibinfo  {journal} {Physics Reports}\ }\textbf {\bibinfo {volume}
  {269}},\ \bibinfo {pages} {133} (\bibinfo {year} {1996})}\BibitemShut
  {NoStop}%
\bibitem [{\citenamefont {Beach}(2004)}]{Beach04}%
  \BibitemOpen
  \bibfield  {author} {\bibinfo {author} {\bibfnamefont {K.~S.~D.}\
  \bibnamefont {Beach}},\ }\href@noop {} {\bibinfo {title} {Identifying the
  maximum entropy method as a special limit of stochastic analytic
  continuation}} (\bibinfo {year} {2004}),\ \Eprint
  {https://arxiv.org/abs/cond-mat/0403055} {arXiv:cond-mat/0403055
  [cond-mat.str-el]} \BibitemShut {NoStop}%
\bibitem [{\citenamefont {Gunnarsson}\ \emph
  {et~al.}(2010{\natexlab{a}})\citenamefont {Gunnarsson}, \citenamefont
  {Haverkort},\ and\ \citenamefont {Sangiovanni}}]{Gunnarsson10}%
  \BibitemOpen
  \bibfield  {author} {\bibinfo {author} {\bibfnamefont {O.}~\bibnamefont
  {Gunnarsson}}, \bibinfo {author} {\bibfnamefont {M.~W.}\ \bibnamefont
  {Haverkort}},\ and\ \bibinfo {author} {\bibfnamefont {G.}~\bibnamefont
  {Sangiovanni}},\ }\href {https://doi.org/10.1103/PhysRevB.82.165125}
  {\bibfield  {journal} {\bibinfo  {journal} {Phys. Rev. B}\ }\textbf {\bibinfo
  {volume} {82}},\ \bibinfo {pages} {165125} (\bibinfo {year}
  {2010}{\natexlab{a}})}\BibitemShut {NoStop}%
\bibitem [{\citenamefont {Gunnarsson}\ \emph
  {et~al.}(2010{\natexlab{b}})\citenamefont {Gunnarsson}, \citenamefont
  {Haverkort},\ and\ \citenamefont {Sangiovanni}}]{Gunnarsson10B}%
  \BibitemOpen
  \bibfield  {author} {\bibinfo {author} {\bibfnamefont {O.}~\bibnamefont
  {Gunnarsson}}, \bibinfo {author} {\bibfnamefont {M.~W.}\ \bibnamefont
  {Haverkort}},\ and\ \bibinfo {author} {\bibfnamefont {G.}~\bibnamefont
  {Sangiovanni}},\ }\href {https://doi.org/10.1103/PhysRevB.81.155107}
  {\bibfield  {journal} {\bibinfo  {journal} {Phys. Rev. B}\ }\textbf {\bibinfo
  {volume} {81}},\ \bibinfo {pages} {155107} (\bibinfo {year}
  {2010}{\natexlab{b}})}\BibitemShut {NoStop}%
\bibitem [{\citenamefont {Bergeron}\ and\ \citenamefont
  {Tremblay}(2016)}]{Bergeron16}%
  \BibitemOpen
  \bibfield  {author} {\bibinfo {author} {\bibfnamefont {D.}~\bibnamefont
  {Bergeron}}\ and\ \bibinfo {author} {\bibfnamefont {A.-M.~S.}\ \bibnamefont
  {Tremblay}},\ }\href {https://doi.org/10.1103/PhysRevE.94.023303} {\bibfield
  {journal} {\bibinfo  {journal} {Phys. Rev. E}\ }\textbf {\bibinfo {volume}
  {94}},\ \bibinfo {pages} {023303} (\bibinfo {year} {2016})}\BibitemShut
  {NoStop}%
\bibitem [{\citenamefont {Levy}\ \emph {et~al.}(2017)\citenamefont {Levy},
  \citenamefont {LeBlanc},\ and\ \citenamefont {Gull}}]{Levy17}%
  \BibitemOpen
  \bibfield  {author} {\bibinfo {author} {\bibfnamefont {R.}~\bibnamefont
  {Levy}}, \bibinfo {author} {\bibfnamefont {J.}~\bibnamefont {LeBlanc}},\ and\
  \bibinfo {author} {\bibfnamefont {E.}~\bibnamefont {Gull}},\ }\href
  {https://doi.org/https://doi.org/10.1016/j.cpc.2017.01.018} {\bibfield
  {journal} {\bibinfo  {journal} {Computer Physics Communications}\ }\textbf
  {\bibinfo {volume} {215}},\ \bibinfo {pages} {149} (\bibinfo {year}
  {2017})}\BibitemShut {NoStop}%
\bibitem [{\citenamefont {Gaenko}\ \emph {et~al.}(2017)\citenamefont {Gaenko},
  \citenamefont {Antipov}, \citenamefont {Carcassi}, \citenamefont {Chen},
  \citenamefont {Chen}, \citenamefont {Dong}, \citenamefont {Gamper},
  \citenamefont {Gukelberger}, \citenamefont {Igarashi}, \citenamefont
  {Iskakov}, \citenamefont {Könz}, \citenamefont {LeBlanc}, \citenamefont
  {Levy}, \citenamefont {Ma}, \citenamefont {Paki}, \citenamefont {Shinaoka},
  \citenamefont {Todo}, \citenamefont {Troyer},\ and\ \citenamefont
  {Gull}}]{Gaenko17}%
  \BibitemOpen
  \bibfield  {author} {\bibinfo {author} {\bibfnamefont {A.}~\bibnamefont
  {Gaenko}}, \bibinfo {author} {\bibfnamefont {A.}~\bibnamefont {Antipov}},
  \bibinfo {author} {\bibfnamefont {G.}~\bibnamefont {Carcassi}}, \bibinfo
  {author} {\bibfnamefont {T.}~\bibnamefont {Chen}}, \bibinfo {author}
  {\bibfnamefont {X.}~\bibnamefont {Chen}}, \bibinfo {author} {\bibfnamefont
  {Q.}~\bibnamefont {Dong}}, \bibinfo {author} {\bibfnamefont {L.}~\bibnamefont
  {Gamper}}, \bibinfo {author} {\bibfnamefont {J.}~\bibnamefont {Gukelberger}},
  \bibinfo {author} {\bibfnamefont {R.}~\bibnamefont {Igarashi}}, \bibinfo
  {author} {\bibfnamefont {S.}~\bibnamefont {Iskakov}}, \bibinfo {author}
  {\bibfnamefont {M.}~\bibnamefont {Könz}}, \bibinfo {author} {\bibfnamefont
  {J.}~\bibnamefont {LeBlanc}}, \bibinfo {author} {\bibfnamefont
  {R.}~\bibnamefont {Levy}}, \bibinfo {author} {\bibfnamefont {P.}~\bibnamefont
  {Ma}}, \bibinfo {author} {\bibfnamefont {J.}~\bibnamefont {Paki}}, \bibinfo
  {author} {\bibfnamefont {H.}~\bibnamefont {Shinaoka}}, \bibinfo {author}
  {\bibfnamefont {S.}~\bibnamefont {Todo}}, \bibinfo {author} {\bibfnamefont
  {M.}~\bibnamefont {Troyer}},\ and\ \bibinfo {author} {\bibfnamefont
  {E.}~\bibnamefont {Gull}},\ }\href
  {https://doi.org/https://doi.org/10.1016/j.cpc.2016.12.009} {\bibfield
  {journal} {\bibinfo  {journal} {Computer Physics Communications}\ }\textbf
  {\bibinfo {volume} {213}},\ \bibinfo {pages} {235 } (\bibinfo {year}
  {2017})}\BibitemShut {NoStop}%
\bibitem [{\citenamefont {Rumetshofer}\ \emph {et~al.}(2019)\citenamefont
  {Rumetshofer}, \citenamefont {Bauernfeind},\ and\ \citenamefont {von~der
  Linden}}]{Rumetshofer19}%
  \BibitemOpen
  \bibfield  {author} {\bibinfo {author} {\bibfnamefont {M.}~\bibnamefont
  {Rumetshofer}}, \bibinfo {author} {\bibfnamefont {D.}~\bibnamefont
  {Bauernfeind}},\ and\ \bibinfo {author} {\bibfnamefont {W.}~\bibnamefont
  {von~der Linden}},\ }\href {https://doi.org/10.1103/PhysRevB.100.075137}
  {\bibfield  {journal} {\bibinfo  {journal} {Phys. Rev. B}\ }\textbf {\bibinfo
  {volume} {100}},\ \bibinfo {pages} {075137} (\bibinfo {year}
  {2019})}\BibitemShut {NoStop}%
\bibitem [{\citenamefont {Sandvik}(1998)}]{Sandvik98}%
  \BibitemOpen
  \bibfield  {author} {\bibinfo {author} {\bibfnamefont {A.~W.}\ \bibnamefont
  {Sandvik}},\ }\href {https://doi.org/10.1103/PhysRevB.57.10287} {\bibfield
  {journal} {\bibinfo  {journal} {Phys. Rev. B}\ }\textbf {\bibinfo {volume}
  {57}},\ \bibinfo {pages} {10287} (\bibinfo {year} {1998})}\BibitemShut
  {NoStop}%
\bibitem [{\citenamefont {Mishchenko}\ \emph {et~al.}(2000)\citenamefont
  {Mishchenko}, \citenamefont {Prokof'ev}, \citenamefont {Sakamoto},\ and\
  \citenamefont {Svistunov}}]{Mishchenko00}%
  \BibitemOpen
  \bibfield  {author} {\bibinfo {author} {\bibfnamefont {A.~S.}\ \bibnamefont
  {Mishchenko}}, \bibinfo {author} {\bibfnamefont {N.~V.}\ \bibnamefont
  {Prokof'ev}}, \bibinfo {author} {\bibfnamefont {A.}~\bibnamefont
  {Sakamoto}},\ and\ \bibinfo {author} {\bibfnamefont {B.~V.}\ \bibnamefont
  {Svistunov}},\ }\href {https://doi.org/10.1103/PhysRevB.62.6317} {\bibfield
  {journal} {\bibinfo  {journal} {Phys. Rev. B}\ }\textbf {\bibinfo {volume}
  {62}},\ \bibinfo {pages} {6317} (\bibinfo {year} {2000})}\BibitemShut
  {NoStop}%
\bibitem [{\citenamefont {Vafayi}\ and\ \citenamefont
  {Gunnarsson}(2007)}]{Gunnarsson07}%
  \BibitemOpen
  \bibfield  {author} {\bibinfo {author} {\bibfnamefont {K.}~\bibnamefont
  {Vafayi}}\ and\ \bibinfo {author} {\bibfnamefont {O.}~\bibnamefont
  {Gunnarsson}},\ }\href {https://doi.org/10.1103/PhysRevB.76.035115}
  {\bibfield  {journal} {\bibinfo  {journal} {Phys. Rev. B}\ }\textbf {\bibinfo
  {volume} {76}},\ \bibinfo {pages} {035115} (\bibinfo {year}
  {2007})}\BibitemShut {NoStop}%
\bibitem [{\citenamefont {Fuchs}\ \emph {et~al.}(2010)\citenamefont {Fuchs},
  \citenamefont {Jarrell},\ and\ \citenamefont {Pruschke}}]{Fuchs10}%
  \BibitemOpen
  \bibfield  {author} {\bibinfo {author} {\bibfnamefont {S.}~\bibnamefont
  {Fuchs}}, \bibinfo {author} {\bibfnamefont {M.}~\bibnamefont {Jarrell}},\
  and\ \bibinfo {author} {\bibfnamefont {T.}~\bibnamefont {Pruschke}},\ }\href
  {https://doi.org/10.1088/1742-6596/200/1/012041} {\bibfield  {journal}
  {\bibinfo  {journal} {Journal of Physics: Conference Series}\ }\textbf
  {\bibinfo {volume} {200}},\ \bibinfo {pages} {012041} (\bibinfo {year}
  {2010})}\BibitemShut {NoStop}%
\bibitem [{\citenamefont {Goulko}\ \emph {et~al.}(2017)\citenamefont {Goulko},
  \citenamefont {Mishchenko}, \citenamefont {Pollet}, \citenamefont
  {Prokof'ev},\ and\ \citenamefont {Svistunov}}]{Goulko17}%
  \BibitemOpen
  \bibfield  {author} {\bibinfo {author} {\bibfnamefont {O.}~\bibnamefont
  {Goulko}}, \bibinfo {author} {\bibfnamefont {A.~S.}\ \bibnamefont
  {Mishchenko}}, \bibinfo {author} {\bibfnamefont {L.}~\bibnamefont {Pollet}},
  \bibinfo {author} {\bibfnamefont {N.}~\bibnamefont {Prokof'ev}},\ and\
  \bibinfo {author} {\bibfnamefont {B.}~\bibnamefont {Svistunov}},\ }\href
  {https://doi.org/10.1103/PhysRevB.95.014102} {\bibfield  {journal} {\bibinfo
  {journal} {Phys. Rev. B}\ }\textbf {\bibinfo {volume} {95}},\ \bibinfo
  {pages} {014102} (\bibinfo {year} {2017})}\BibitemShut {NoStop}%
\bibitem [{\citenamefont {Krivenko}\ and\ \citenamefont
  {Harland}(2019)}]{Krivenko19}%
  \BibitemOpen
  \bibfield  {author} {\bibinfo {author} {\bibfnamefont {I.}~\bibnamefont
  {Krivenko}}\ and\ \bibinfo {author} {\bibfnamefont {M.}~\bibnamefont
  {Harland}},\ }\href
  {https://doi.org/https://doi.org/10.1016/j.cpc.2019.01.021} {\bibfield
  {journal} {\bibinfo  {journal} {Computer Physics Communications}\ }\textbf
  {\bibinfo {volume} {239}},\ \bibinfo {pages} {166 } (\bibinfo {year}
  {2019})}\BibitemShut {NoStop}%
\bibitem [{\citenamefont {Otsuki}\ \emph {et~al.}(2017)\citenamefont {Otsuki},
  \citenamefont {Ohzeki}, \citenamefont {Shinaoka},\ and\ \citenamefont
  {Yoshimi}}]{Otsuki17}%
  \BibitemOpen
  \bibfield  {author} {\bibinfo {author} {\bibfnamefont {J.}~\bibnamefont
  {Otsuki}}, \bibinfo {author} {\bibfnamefont {M.}~\bibnamefont {Ohzeki}},
  \bibinfo {author} {\bibfnamefont {H.}~\bibnamefont {Shinaoka}},\ and\
  \bibinfo {author} {\bibfnamefont {K.}~\bibnamefont {Yoshimi}},\ }\href
  {https://doi.org/10.1103/PhysRevE.95.061302} {\bibfield  {journal} {\bibinfo
  {journal} {Phys. Rev. E}\ }\textbf {\bibinfo {volume} {95}},\ \bibinfo
  {pages} {061302} (\bibinfo {year} {2017})}\BibitemShut {NoStop}%
\bibitem [{\citenamefont {Otsuki}\ \emph {et~al.}(2020)\citenamefont {Otsuki},
  \citenamefont {Ohzeki}, \citenamefont {Shinaoka},\ and\ \citenamefont
  {Yoshimi}}]{Otsuki20}%
  \BibitemOpen
  \bibfield  {author} {\bibinfo {author} {\bibfnamefont {J.}~\bibnamefont
  {Otsuki}}, \bibinfo {author} {\bibfnamefont {M.}~\bibnamefont {Ohzeki}},
  \bibinfo {author} {\bibfnamefont {H.}~\bibnamefont {Shinaoka}},\ and\
  \bibinfo {author} {\bibfnamefont {K.}~\bibnamefont {Yoshimi}},\ }\href
  {https://doi.org/10.7566/JPSJ.89.012001} {\bibfield  {journal} {\bibinfo
  {journal} {Journal of the Physical Society of Japan}\ }\textbf {\bibinfo
  {volume} {89}},\ \bibinfo {pages} {012001} (\bibinfo {year}
  {2020})}\BibitemShut {NoStop}%
\bibitem [{\citenamefont {Yoon}\ \emph {et~al.}(2018)\citenamefont {Yoon},
  \citenamefont {Sim},\ and\ \citenamefont {Han}}]{Yoon18}%
  \BibitemOpen
  \bibfield  {author} {\bibinfo {author} {\bibfnamefont {H.}~\bibnamefont
  {Yoon}}, \bibinfo {author} {\bibfnamefont {J.-H.}\ \bibnamefont {Sim}},\ and\
  \bibinfo {author} {\bibfnamefont {M.~J.}\ \bibnamefont {Han}},\ }\href
  {https://doi.org/10.1103/PhysRevB.98.245101} {\bibfield  {journal} {\bibinfo
  {journal} {Phys. Rev. B}\ }\textbf {\bibinfo {volume} {98}},\ \bibinfo
  {pages} {245101} (\bibinfo {year} {2018})}\BibitemShut {NoStop}%
\bibitem [{\citenamefont {Vidberg}\ and\ \citenamefont
  {Serene}(1977)}]{Vidberg77}%
  \BibitemOpen
  \bibfield  {author} {\bibinfo {author} {\bibfnamefont {H.~J.}\ \bibnamefont
  {Vidberg}}\ and\ \bibinfo {author} {\bibfnamefont {J.~W.}\ \bibnamefont
  {Serene}},\ }\href {https://doi.org/10.1007/BF00655090} {\bibfield  {journal}
  {\bibinfo  {journal} {Journal of Low Temperature Physics}\ }\textbf {\bibinfo
  {volume} {29}},\ \bibinfo {pages} {179} (\bibinfo {year} {1977})}\BibitemShut
  {NoStop}%
\bibitem [{\citenamefont {Beach}\ \emph {et~al.}(2000)\citenamefont {Beach},
  \citenamefont {Gooding},\ and\ \citenamefont {Marsiglio}}]{Beach00}%
  \BibitemOpen
  \bibfield  {author} {\bibinfo {author} {\bibfnamefont {K.~S.~D.}\
  \bibnamefont {Beach}}, \bibinfo {author} {\bibfnamefont {R.~J.}\ \bibnamefont
  {Gooding}},\ and\ \bibinfo {author} {\bibfnamefont {F.}~\bibnamefont
  {Marsiglio}},\ }\href {https://doi.org/10.1103/PhysRevB.61.5147} {\bibfield
  {journal} {\bibinfo  {journal} {Phys. Rev. B}\ }\textbf {\bibinfo {volume}
  {61}},\ \bibinfo {pages} {5147} (\bibinfo {year} {2000})}\BibitemShut
  {NoStop}%
\bibitem [{\citenamefont {\"Ostlin}\ \emph {et~al.}(2012)\citenamefont
  {\"Ostlin}, \citenamefont {Chioncel},\ and\ \citenamefont
  {Vitos}}]{Ostlin12}%
  \BibitemOpen
  \bibfield  {author} {\bibinfo {author} {\bibfnamefont {A.}~\bibnamefont
  {\"Ostlin}}, \bibinfo {author} {\bibfnamefont {L.}~\bibnamefont {Chioncel}},\
  and\ \bibinfo {author} {\bibfnamefont {L.}~\bibnamefont {Vitos}},\ }\href
  {https://doi.org/10.1103/PhysRevB.86.235107} {\bibfield  {journal} {\bibinfo
  {journal} {Phys. Rev. B}\ }\textbf {\bibinfo {volume} {86}},\ \bibinfo
  {pages} {235107} (\bibinfo {year} {2012})}\BibitemShut {NoStop}%
\bibitem [{\citenamefont {{Osolin, {\ifmmode \check{Z}\else \v{Z}\fi{}iga}}}\
  and\ \citenamefont {\ifmmode~\check{Z}\else
  \v{Z}\fi{}itko}(2013)}]{Osolin13}%
  \BibitemOpen
  \bibfield  {author} {\bibinfo {author} {\bibnamefont {{Osolin, {\ifmmode
  \check{Z}\else \v{Z}\fi{}iga}}}}\ and\ \bibinfo {author} {\bibfnamefont
  {R.}~\bibnamefont {\ifmmode~\check{Z}\else \v{Z}\fi{}itko}},\ }\href
  {https://doi.org/10.1103/PhysRevB.87.245135} {\bibfield  {journal} {\bibinfo
  {journal} {Phys. Rev. B}\ }\textbf {\bibinfo {volume} {87}},\ \bibinfo
  {pages} {245135} (\bibinfo {year} {2013})}\BibitemShut {NoStop}%
\bibitem [{\citenamefont {Sch\"ott}\ \emph {et~al.}(2016)\citenamefont
  {Sch\"ott}, \citenamefont {Locht}, \citenamefont {Lundin}, \citenamefont
  {Gr\aa{}n\"as}, \citenamefont {Eriksson},\ and\ \citenamefont
  {Di~Marco}}]{Schott16}%
  \BibitemOpen
  \bibfield  {author} {\bibinfo {author} {\bibfnamefont {J.}~\bibnamefont
  {Sch\"ott}}, \bibinfo {author} {\bibfnamefont {I.~L.~M.}\ \bibnamefont
  {Locht}}, \bibinfo {author} {\bibfnamefont {E.}~\bibnamefont {Lundin}},
  \bibinfo {author} {\bibfnamefont {O.}~\bibnamefont {Gr\aa{}n\"as}}, \bibinfo
  {author} {\bibfnamefont {O.}~\bibnamefont {Eriksson}},\ and\ \bibinfo
  {author} {\bibfnamefont {I.}~\bibnamefont {Di~Marco}},\ }\href
  {https://doi.org/10.1103/PhysRevB.93.075104} {\bibfield  {journal} {\bibinfo
  {journal} {Phys. Rev. B}\ }\textbf {\bibinfo {volume} {93}},\ \bibinfo
  {pages} {075104} (\bibinfo {year} {2016})}\BibitemShut {NoStop}%
\bibitem [{\citenamefont {Han}\ \emph {et~al.}(2017)\citenamefont {Han},
  \citenamefont {Liao}, \citenamefont {Xie}, \citenamefont {Huang},
  \citenamefont {Meng},\ and\ \citenamefont {Xiang}}]{Han17}%
  \BibitemOpen
  \bibfield  {author} {\bibinfo {author} {\bibfnamefont {X.-J.}\ \bibnamefont
  {Han}}, \bibinfo {author} {\bibfnamefont {H.-J.}\ \bibnamefont {Liao}},
  \bibinfo {author} {\bibfnamefont {H.-D.}\ \bibnamefont {Xie}}, \bibinfo
  {author} {\bibfnamefont {R.-Z.}\ \bibnamefont {Huang}}, \bibinfo {author}
  {\bibfnamefont {Z.-Y.}\ \bibnamefont {Meng}},\ and\ \bibinfo {author}
  {\bibfnamefont {T.}~\bibnamefont {Xiang}},\ }\href
  {https://doi.org/10.1088/0256-307x/34/7/077102} {\bibfield  {journal}
  {\bibinfo  {journal} {Chinese Physics Letters}\ }\textbf {\bibinfo {volume}
  {34}},\ \bibinfo {pages} {077102} (\bibinfo {year} {2017})}\BibitemShut
  {NoStop}%
\bibitem [{\citenamefont {Tomczak}\ and\ \citenamefont
  {Biermann}(2007)}]{Tomczak07}%
  \BibitemOpen
  \bibfield  {author} {\bibinfo {author} {\bibfnamefont {J.~M.}\ \bibnamefont
  {Tomczak}}\ and\ \bibinfo {author} {\bibfnamefont {S.}~\bibnamefont
  {Biermann}},\ }\href {https://doi.org/10.1088/0953-8984/19/36/365206}
  {\bibfield  {journal} {\bibinfo  {journal} {Journal of Physics: Condensed
  Matter}\ }\textbf {\bibinfo {volume} {19}},\ \bibinfo {pages} {365206}
  (\bibinfo {year} {2007})}\BibitemShut {NoStop}%
\bibitem [{\citenamefont {Dang}\ \emph {et~al.}(2014)\citenamefont {Dang},
  \citenamefont {Ai}, \citenamefont {Millis},\ and\ \citenamefont
  {Marianetti}}]{Dang14}%
  \BibitemOpen
  \bibfield  {author} {\bibinfo {author} {\bibfnamefont {H.~T.}\ \bibnamefont
  {Dang}}, \bibinfo {author} {\bibfnamefont {X.}~\bibnamefont {Ai}}, \bibinfo
  {author} {\bibfnamefont {A.~J.}\ \bibnamefont {Millis}},\ and\ \bibinfo
  {author} {\bibfnamefont {C.~A.}\ \bibnamefont {Marianetti}},\ }\href
  {https://doi.org/10.1103/PhysRevB.90.125114} {\bibfield  {journal} {\bibinfo
  {journal} {Phys. Rev. B}\ }\textbf {\bibinfo {volume} {90}},\ \bibinfo
  {pages} {125114} (\bibinfo {year} {2014})}\BibitemShut {NoStop}%
\bibitem [{\citenamefont {Gull}\ and\ \citenamefont {Millis}(2014)}]{Gull14}%
  \BibitemOpen
  \bibfield  {author} {\bibinfo {author} {\bibfnamefont {E.}~\bibnamefont
  {Gull}}\ and\ \bibinfo {author} {\bibfnamefont {A.~J.}\ \bibnamefont
  {Millis}},\ }\href {https://doi.org/10.1103/PhysRevB.90.041110} {\bibfield
  {journal} {\bibinfo  {journal} {Phys. Rev. B}\ }\textbf {\bibinfo {volume}
  {90}},\ \bibinfo {pages} {041110} (\bibinfo {year} {2014})}\BibitemShut
  {NoStop}%
\bibitem [{\citenamefont {Kraberger}\ \emph {et~al.}(2017)\citenamefont
  {Kraberger}, \citenamefont {Triebl}, \citenamefont {Zingl},\ and\
  \citenamefont {Aichhorn}}]{Kraberger17}%
  \BibitemOpen
  \bibfield  {author} {\bibinfo {author} {\bibfnamefont {G.~J.}\ \bibnamefont
  {Kraberger}}, \bibinfo {author} {\bibfnamefont {R.}~\bibnamefont {Triebl}},
  \bibinfo {author} {\bibfnamefont {M.}~\bibnamefont {Zingl}},\ and\ \bibinfo
  {author} {\bibfnamefont {M.}~\bibnamefont {Aichhorn}},\ }\href
  {https://doi.org/10.1103/PhysRevB.96.155128} {\bibfield  {journal} {\bibinfo
  {journal} {Phys. Rev. B}\ }\textbf {\bibinfo {volume} {96}},\ \bibinfo
  {pages} {155128} (\bibinfo {year} {2017})}\BibitemShut {NoStop}%
\bibitem [{\citenamefont {Sim}\ and\ \citenamefont {Han}(2018)}]{Sim18}%
  \BibitemOpen
  \bibfield  {author} {\bibinfo {author} {\bibfnamefont {J.-H.}\ \bibnamefont
  {Sim}}\ and\ \bibinfo {author} {\bibfnamefont {M.~J.}\ \bibnamefont {Han}},\
  }\href {https://doi.org/10.1103/PhysRevB.98.205102} {\bibfield  {journal}
  {\bibinfo  {journal} {Phys. Rev. B}\ }\textbf {\bibinfo {volume} {98}},\
  \bibinfo {pages} {205102} (\bibinfo {year} {2018})}\BibitemShut {NoStop}%
\bibitem [{\citenamefont {Fei}\ \emph {et~al.}(2021)\citenamefont {Fei},
  \citenamefont {Yeh},\ and\ \citenamefont {Gull}}]{Fei21}%
  \BibitemOpen
  \bibfield  {author} {\bibinfo {author} {\bibfnamefont {J.}~\bibnamefont
  {Fei}}, \bibinfo {author} {\bibfnamefont {C.-N.}\ \bibnamefont {Yeh}},\ and\
  \bibinfo {author} {\bibfnamefont {E.}~\bibnamefont {Gull}},\ }\href
  {https://doi.org/10.1103/PhysRevLett.126.056402} {\bibfield  {journal}
  {\bibinfo  {journal} {Phys. Rev. Lett.}\ }\textbf {\bibinfo {volume} {126}},\
  \bibinfo {pages} {056402} (\bibinfo {year} {2021})}\BibitemShut {NoStop}%
\bibitem [{\citenamefont {Stanescu}\ and\ \citenamefont
  {Kotliar}(2006)}]{Tudor06}%
  \BibitemOpen
  \bibfield  {author} {\bibinfo {author} {\bibfnamefont {T.~D.}\ \bibnamefont
  {Stanescu}}\ and\ \bibinfo {author} {\bibfnamefont {G.}~\bibnamefont
  {Kotliar}},\ }\href {https://doi.org/10.1103/PhysRevB.74.125110} {\bibfield
  {journal} {\bibinfo  {journal} {Phys. Rev. B}\ }\textbf {\bibinfo {volume}
  {74}},\ \bibinfo {pages} {125110} (\bibinfo {year} {2006})}\BibitemShut
  {NoStop}%
\bibitem [{\citenamefont {Carathéodory}(1907)}]{Caratheodory07}%
  \BibitemOpen
  \bibfield  {author} {\bibinfo {author} {\bibfnamefont {C.}~\bibnamefont
  {Carathéodory}},\ }\href
  {https://doi.org/https://doi.org/10.1007/BF01449883} {\bibfield  {journal}
  {\bibinfo  {journal} {Mathematische Annalen}\ }\textbf {\bibinfo {volume}
  {64}},\ \bibinfo {pages} {95} (\bibinfo {year} {1907})}\BibitemShut {NoStop}%
\bibitem [{\citenamefont {Schur}(1918)}]{Schur18}%
  \BibitemOpen
  \bibfield  {author} {\bibinfo {author} {\bibfnamefont {J.}~\bibnamefont
  {Schur}},\ }\href {http://eudml.org/doc/149476} {\bibfield  {journal}
  {\bibinfo  {journal} {Journal für die reine und angewandte Mathematik}\
  }\textbf {\bibinfo {volume} {148}},\ \bibinfo {pages} {122} (\bibinfo {year}
  {1918})}\BibitemShut {NoStop}%
\bibitem [{\citenamefont {Delsarte}\ \emph {et~al.}(1979)\citenamefont
  {Delsarte}, \citenamefont {Genin},\ and\ \citenamefont {Kamp}}]{Kamp79}%
  \BibitemOpen
  \bibfield  {author} {\bibinfo {author} {\bibfnamefont {P.}~\bibnamefont
  {Delsarte}}, \bibinfo {author} {\bibfnamefont {Y.}~\bibnamefont {Genin}},\
  and\ \bibinfo {author} {\bibfnamefont {Y.}~\bibnamefont {Kamp}},\ }\href
  {http://www.jstor.org/stable/2100767} {\bibfield  {journal} {\bibinfo
  {journal} {SIAM Journal on Applied Mathematics}\ }\textbf {\bibinfo {volume}
  {36}},\ \bibinfo {pages} {47} (\bibinfo {year} {1979})}\BibitemShut {NoStop}%
\bibitem [{\citenamefont {Arov}\ and\ \citenamefont {Rozhenko}(2008)}]{Arov08}%
  \BibitemOpen
  \bibfield  {author} {\bibinfo {author} {\bibfnamefont {D.}~\bibnamefont
  {Arov}}\ and\ \bibinfo {author} {\bibfnamefont {N.}~\bibnamefont
  {Rozhenko}},\ }\href {https://doi.org/10.1090/S1061-0022-08-01002-9}
  {\bibfield  {journal} {\bibinfo  {journal} {St Petersburg Mathematical
  Journal}\ }\textbf {\bibinfo {volume} {19}},\ \bibinfo {pages} {375}
  (\bibinfo {year} {2008})}\BibitemShut {NoStop}%
\bibitem [{\citenamefont {Fritzsche}\ \emph {et~al.}(2012)\citenamefont
  {Fritzsche}, \citenamefont {Kirstein}, \citenamefont {Lasarow},\ and\
  \citenamefont {Rahn}}]{Fritzsche12}%
  \BibitemOpen
  \bibfield  {author} {\bibinfo {author} {\bibfnamefont {B.}~\bibnamefont
  {Fritzsche}}, \bibinfo {author} {\bibfnamefont {B.}~\bibnamefont {Kirstein}},
  \bibinfo {author} {\bibfnamefont {A.}~\bibnamefont {Lasarow}},\ and\ \bibinfo
  {author} {\bibfnamefont {A.}~\bibnamefont {Rahn}},\ }\bibinfo {title} {On
  reciprocal sequences of matricial carath{\'e}odory sequences and associated
  matrix functions},\ in\ \href {https://doi.org/10.1007/978-3-0348-0428-8_2}
  {\emph {\bibinfo {booktitle} {Interpolation, Schur Functions and Moment
  Problems II}}},\ \bibinfo {editor} {edited by\ \bibinfo {editor}
  {\bibfnamefont {D.}~\bibnamefont {Alpay}}\ and\ \bibinfo {editor}
  {\bibfnamefont {B.}~\bibnamefont {Kirstein}}}\ (\bibinfo  {publisher}
  {Springer Basel},\ \bibinfo {address} {Basel},\ \bibinfo {year} {2012})\ pp.\
  \bibinfo {pages} {57--115}\BibitemShut {NoStop}%
\bibitem [{\citenamefont {Pick}(1917)}]{Pick}%
  \BibitemOpen
  \bibfield  {author} {\bibinfo {author} {\bibfnamefont {G.}~\bibnamefont
  {Pick}},\ }\href {https://doi.org/10.1007/BF01457103} {\bibfield  {journal}
  {\bibinfo  {journal} {Mathematische annalen}\ }\textbf {\bibinfo {volume}
  {78}},\ \bibinfo {pages} {270 } (\bibinfo {year} {1917})}\BibitemShut
  {NoStop}%
\bibitem [{\citenamefont {Akhiezer}(1965)}]{Akhiezer}%
  \BibitemOpen
  \bibfield  {author} {\bibinfo {author} {\bibfnamefont {N.}~\bibnamefont
  {Akhiezer}},\ }\href@noop {} {\emph {\bibinfo {title} {The Classical Moment
  Problem and Some Related Questions in Analysis}}}\ (\bibinfo  {publisher}
  {Oliver and Boyd.},\ \bibinfo {address} {London},\ \bibinfo {year}
  {1965})\BibitemShut {NoStop}%
\bibitem [{\citenamefont {Khargonekar}\ and\ \citenamefont
  {Tannenbaum}(1917)}]{Tannenbaum17}%
  \BibitemOpen
  \bibfield  {author} {\bibinfo {author} {\bibfnamefont {P.}~\bibnamefont
  {Khargonekar}}\ and\ \bibinfo {author} {\bibfnamefont {A.}~\bibnamefont
  {Tannenbaum}},\ }\href {https://doi.org/10.1109/TAC.1985.1103805} {\bibfield
  {journal} {\bibinfo  {journal} {IEEE Transactions on Automatic Control}\
  }\textbf {\bibinfo {volume} {30}},\ \bibinfo {pages} {1005 } (\bibinfo {year}
  {1917})}\BibitemShut {NoStop}%
\bibitem [{\citenamefont {Chen}\ and\ \citenamefont {Çetin
  Kaya~Koç}(1994)}]{Chen94}%
  \BibitemOpen
  \bibfield  {author} {\bibinfo {author} {\bibfnamefont {G.}~\bibnamefont
  {Chen}}\ and\ \bibinfo {author} {\bibnamefont {Çetin Kaya~Koç}},\ }\href
  {https://doi.org/https://doi.org/10.1016/0024-3795(94)90205-4} {\bibfield
  {journal} {\bibinfo  {journal} {Linear Algebra and its Applications}\
  }\textbf {\bibinfo {volume} {203-204}},\ \bibinfo {pages} {253 } (\bibinfo
  {year} {1994})}\BibitemShut {NoStop}%
\bibitem [{\citenamefont {Yazici}\ and\ \citenamefont
  {Sevindir}(2013)}]{Cuneyt13}%
  \BibitemOpen
  \bibfield  {author} {\bibinfo {author} {\bibfnamefont {C.}~\bibnamefont
  {Yazici}}\ and\ \bibinfo {author} {\bibfnamefont {H.~K.}\ \bibnamefont
  {Sevindir}},\ }\href {https://doi.org/10.1063/1.4826042} {\bibfield
  {journal} {\bibinfo  {journal} {AIP Conference Proceedings}\ }\textbf
  {\bibinfo {volume} {1558}},\ \bibinfo {pages} {2474} (\bibinfo {year}
  {2013})}\BibitemShut {NoStop}%
\bibitem [{\citenamefont {Potapov}(1955)}]{Potapov55}%
  \BibitemOpen
  \bibfield  {author} {\bibinfo {author} {\bibfnamefont {V.}~\bibnamefont
  {Potapov}},\ }\href@noop {} {\emph {\bibinfo {title} {The multiplicative
  structure of J-contractive matrix functions}}}\ (\bibinfo  {publisher}
  {GITTL},\ \bibinfo {address} {Moscow},\ \bibinfo {year} {1955})\ pp.\
  \bibinfo {pages} {125 -- 236}\BibitemShut {NoStop}%
\bibitem [{\citenamefont {Balzer}\ and\ \citenamefont
  {Eckstein}(2014)}]{Karsten14}%
  \BibitemOpen
  \bibfield  {author} {\bibinfo {author} {\bibfnamefont {K.}~\bibnamefont
  {Balzer}}\ and\ \bibinfo {author} {\bibfnamefont {M.}~\bibnamefont
  {Eckstein}},\ }\href {https://doi.org/10.1103/PhysRevB.89.035148} {\bibfield
  {journal} {\bibinfo  {journal} {Phys. Rev. B}\ }\textbf {\bibinfo {volume}
  {89}},\ \bibinfo {pages} {035148} (\bibinfo {year} {2014})}\BibitemShut
  {NoStop}%
\bibitem [{\citenamefont {Gramsch}\ and\ \citenamefont
  {Potthoff}(2015)}]{Christian15}%
  \BibitemOpen
  \bibfield  {author} {\bibinfo {author} {\bibfnamefont {C.}~\bibnamefont
  {Gramsch}}\ and\ \bibinfo {author} {\bibfnamefont {M.}~\bibnamefont
  {Potthoff}},\ }\href {https://doi.org/10.1103/PhysRevB.92.235135} {\bibfield
  {journal} {\bibinfo  {journal} {Phys. Rev. B}\ }\textbf {\bibinfo {volume}
  {92}},\ \bibinfo {pages} {235135} (\bibinfo {year} {2015})}\BibitemShut
  {NoStop}%
\bibitem [{\citenamefont {Sakai}\ \emph {et~al.}(2012)\citenamefont {Sakai},
  \citenamefont {Sangiovanni}, \citenamefont {Civelli}, \citenamefont {Motome},
  \citenamefont {Held},\ and\ \citenamefont {Imada}}]{Shiro12}%
  \BibitemOpen
  \bibfield  {author} {\bibinfo {author} {\bibfnamefont {S.}~\bibnamefont
  {Sakai}}, \bibinfo {author} {\bibfnamefont {G.}~\bibnamefont {Sangiovanni}},
  \bibinfo {author} {\bibfnamefont {M.}~\bibnamefont {Civelli}}, \bibinfo
  {author} {\bibfnamefont {Y.}~\bibnamefont {Motome}}, \bibinfo {author}
  {\bibfnamefont {K.}~\bibnamefont {Held}},\ and\ \bibinfo {author}
  {\bibfnamefont {M.}~\bibnamefont {Imada}},\ }\href
  {https://doi.org/10.1103/PhysRevB.85.035102} {\bibfield  {journal} {\bibinfo
  {journal} {Phys. Rev. B}\ }\textbf {\bibinfo {volume} {85}},\ \bibinfo
  {pages} {035102} (\bibinfo {year} {2012})}\BibitemShut {NoStop}%
\bibitem [{\citenamefont {Johnson}(1972)}]{Johnson72}%
  \BibitemOpen
  \bibfield  {author} {\bibinfo {author} {\bibfnamefont {C.~R.}\ \bibnamefont
  {Johnson}},\ }\emph {\bibinfo {title} {Matrices whose hermitian part is
  positive definite}},\ \href {https://doi.org/10.7907/ZXNF-SB10} {Ph.D.
  thesis} (\bibinfo {year} {1972})\BibitemShut {NoStop}%
\bibitem [{\citenamefont {Qin}\ \emph {et~al.}(2021)\citenamefont {Qin},
  \citenamefont {Schäfer}, \citenamefont {Andergassen}, \citenamefont
  {Corboz},\ and\ \citenamefont {Gull}}]{Qin21}%
  \BibitemOpen
  \bibfield  {author} {\bibinfo {author} {\bibfnamefont {M.}~\bibnamefont
  {Qin}}, \bibinfo {author} {\bibfnamefont {T.}~\bibnamefont {Schäfer}},
  \bibinfo {author} {\bibfnamefont {S.}~\bibnamefont {Andergassen}}, \bibinfo
  {author} {\bibfnamefont {P.}~\bibnamefont {Corboz}},\ and\ \bibinfo {author}
  {\bibfnamefont {E.}~\bibnamefont {Gull}},\ }\href@noop {} {\bibinfo {title}
  {The hubbard model: A computational perspective}} (\bibinfo {year} {2021}),\
  \Eprint {https://arxiv.org/abs/2104.00064} {arXiv:2104.00064
  [cond-mat.str-el]} \BibitemShut {NoStop}%
\bibitem [{\citenamefont {Shinaoka}\ \emph {et~al.}(2017)\citenamefont
  {Shinaoka}, \citenamefont {Otsuki}, \citenamefont {Ohzeki},\ and\
  \citenamefont {Yoshimi}}]{Shinaoka17}%
  \BibitemOpen
  \bibfield  {author} {\bibinfo {author} {\bibfnamefont {H.}~\bibnamefont
  {Shinaoka}}, \bibinfo {author} {\bibfnamefont {J.}~\bibnamefont {Otsuki}},
  \bibinfo {author} {\bibfnamefont {M.}~\bibnamefont {Ohzeki}},\ and\ \bibinfo
  {author} {\bibfnamefont {K.}~\bibnamefont {Yoshimi}},\ }\href
  {https://doi.org/10.1103/PhysRevB.96.035147} {\bibfield  {journal} {\bibinfo
  {journal} {Phys. Rev. B}\ }\textbf {\bibinfo {volume} {96}},\ \bibinfo
  {pages} {035147} (\bibinfo {year} {2017})}\BibitemShut {NoStop}%
\bibitem [{\citenamefont {Li}\ \emph {et~al.}(2020)\citenamefont {Li},
  \citenamefont {Wallerberger}, \citenamefont {Chikano}, \citenamefont {Yeh},
  \citenamefont {Gull},\ and\ \citenamefont {Shinaoka}}]{Li20}%
  \BibitemOpen
  \bibfield  {author} {\bibinfo {author} {\bibfnamefont {J.}~\bibnamefont
  {Li}}, \bibinfo {author} {\bibfnamefont {M.}~\bibnamefont {Wallerberger}},
  \bibinfo {author} {\bibfnamefont {N.}~\bibnamefont {Chikano}}, \bibinfo
  {author} {\bibfnamefont {C.-N.}\ \bibnamefont {Yeh}}, \bibinfo {author}
  {\bibfnamefont {E.}~\bibnamefont {Gull}},\ and\ \bibinfo {author}
  {\bibfnamefont {H.}~\bibnamefont {Shinaoka}},\ }\href
  {https://doi.org/10.1103/PhysRevB.101.035144} {\bibfield  {journal} {\bibinfo
   {journal} {Phys. Rev. B}\ }\textbf {\bibinfo {volume} {101}},\ \bibinfo
  {pages} {035144} (\bibinfo {year} {2020})}\BibitemShut {NoStop}%
\bibitem [{\citenamefont {Wang}\ \emph {et~al.}(2009)\citenamefont {Wang},
  \citenamefont {Gull}, \citenamefont {de' Medici}, \citenamefont {Capone},\
  and\ \citenamefont {Millis}}]{Wang09}%
  \BibitemOpen
  \bibfield  {author} {\bibinfo {author} {\bibfnamefont {X.}~\bibnamefont
  {Wang}}, \bibinfo {author} {\bibfnamefont {E.}~\bibnamefont {Gull}}, \bibinfo
  {author} {\bibfnamefont {L.}~\bibnamefont {de' Medici}}, \bibinfo {author}
  {\bibfnamefont {M.}~\bibnamefont {Capone}},\ and\ \bibinfo {author}
  {\bibfnamefont {A.~J.}\ \bibnamefont {Millis}},\ }\href
  {https://doi.org/10.1103/PhysRevB.80.045101} {\bibfield  {journal} {\bibinfo
  {journal} {Phys. Rev. B}\ }\textbf {\bibinfo {volume} {80}},\ \bibinfo
  {pages} {045101} (\bibinfo {year} {2009})}\BibitemShut {NoStop}%
\bibitem [{\citenamefont {VandeVondele}\ and\ \citenamefont
  {Hutter}(2007)}]{VandeVondele07}%
  \BibitemOpen
  \bibfield  {author} {\bibinfo {author} {\bibfnamefont {J.}~\bibnamefont
  {VandeVondele}}\ and\ \bibinfo {author} {\bibfnamefont {J.}~\bibnamefont
  {Hutter}},\ }\href {https://doi.org/10.1063/1.2770708} {\bibfield  {journal}
  {\bibinfo  {journal} {The Journal of Chemical Physics}\ }\textbf {\bibinfo
  {volume} {127}},\ \bibinfo {pages} {114105} (\bibinfo {year}
  {2007})}\BibitemShut {NoStop}%
\bibitem [{\citenamefont {Goedecker}\ \emph {et~al.}(1996)\citenamefont
  {Goedecker}, \citenamefont {Teter},\ and\ \citenamefont
  {Hutter}}]{Goedecker96}%
  \BibitemOpen
  \bibfield  {author} {\bibinfo {author} {\bibfnamefont {S.}~\bibnamefont
  {Goedecker}}, \bibinfo {author} {\bibfnamefont {M.}~\bibnamefont {Teter}},\
  and\ \bibinfo {author} {\bibfnamefont {J.}~\bibnamefont {Hutter}},\ }\href
  {https://doi.org/10.1103/PhysRevB.54.1703} {\bibfield  {journal} {\bibinfo
  {journal} {Phys. Rev. B}\ }\textbf {\bibinfo {volume} {54}},\ \bibinfo
  {pages} {1703} (\bibinfo {year} {1996})}\BibitemShut {NoStop}%
\bibitem [{\citenamefont {Sun}\ \emph {et~al.}(2020)\citenamefont {Sun},
  \citenamefont {Zhang}, \citenamefont {Banerjee}, \citenamefont {Bao},
  \citenamefont {Barbry}, \citenamefont {Blunt}, \citenamefont {Bogdanov},
  \citenamefont {Booth}, \citenamefont {Chen}, \citenamefont {Cui},
  \citenamefont {Eriksen}, \citenamefont {Gao}, \citenamefont {Guo},
  \citenamefont {Hermann}, \citenamefont {Hermes}, \citenamefont {Koh},
  \citenamefont {Koval}, \citenamefont {Lehtola}, \citenamefont {Li},
  \citenamefont {Liu}, \citenamefont {Mardirossian}, \citenamefont {McClain},
  \citenamefont {Motta}, \citenamefont {Mussard}, \citenamefont {Pham},
  \citenamefont {Pulkin}, \citenamefont {Purwanto}, \citenamefont {Robinson},
  \citenamefont {Ronca}, \citenamefont {Sayfutyarova}, \citenamefont
  {Scheurer}, \citenamefont {Schurkus}, \citenamefont {Smith}, \citenamefont
  {Sun}, \citenamefont {Sun}, \citenamefont {Upadhyay}, \citenamefont {Wagner},
  \citenamefont {Wang}, \citenamefont {White}, \citenamefont {Whitfield},
  \citenamefont {Williamson}, \citenamefont {Wouters}, \citenamefont {Yang},
  \citenamefont {Yu}, \citenamefont {Zhu}, \citenamefont {Berkelbach},
  \citenamefont {Sharma}, \citenamefont {Sokolov},\ and\ \citenamefont
  {Chan}}]{pySCF20}%
  \BibitemOpen
  \bibfield  {author} {\bibinfo {author} {\bibfnamefont {Q.}~\bibnamefont
  {Sun}}, \bibinfo {author} {\bibfnamefont {X.}~\bibnamefont {Zhang}}, \bibinfo
  {author} {\bibfnamefont {S.}~\bibnamefont {Banerjee}}, \bibinfo {author}
  {\bibfnamefont {P.}~\bibnamefont {Bao}}, \bibinfo {author} {\bibfnamefont
  {M.}~\bibnamefont {Barbry}}, \bibinfo {author} {\bibfnamefont {N.~S.}\
  \bibnamefont {Blunt}}, \bibinfo {author} {\bibfnamefont {N.~A.}\ \bibnamefont
  {Bogdanov}}, \bibinfo {author} {\bibfnamefont {G.~H.}\ \bibnamefont {Booth}},
  \bibinfo {author} {\bibfnamefont {J.}~\bibnamefont {Chen}}, \bibinfo {author}
  {\bibfnamefont {Z.-H.}\ \bibnamefont {Cui}}, \bibinfo {author} {\bibfnamefont
  {J.~J.}\ \bibnamefont {Eriksen}}, \bibinfo {author} {\bibfnamefont
  {Y.}~\bibnamefont {Gao}}, \bibinfo {author} {\bibfnamefont {S.}~\bibnamefont
  {Guo}}, \bibinfo {author} {\bibfnamefont {J.}~\bibnamefont {Hermann}},
  \bibinfo {author} {\bibfnamefont {M.~R.}\ \bibnamefont {Hermes}}, \bibinfo
  {author} {\bibfnamefont {K.}~\bibnamefont {Koh}}, \bibinfo {author}
  {\bibfnamefont {P.}~\bibnamefont {Koval}}, \bibinfo {author} {\bibfnamefont
  {S.}~\bibnamefont {Lehtola}}, \bibinfo {author} {\bibfnamefont
  {Z.}~\bibnamefont {Li}}, \bibinfo {author} {\bibfnamefont {J.}~\bibnamefont
  {Liu}}, \bibinfo {author} {\bibfnamefont {N.}~\bibnamefont {Mardirossian}},
  \bibinfo {author} {\bibfnamefont {J.~D.}\ \bibnamefont {McClain}}, \bibinfo
  {author} {\bibfnamefont {M.}~\bibnamefont {Motta}}, \bibinfo {author}
  {\bibfnamefont {B.}~\bibnamefont {Mussard}}, \bibinfo {author} {\bibfnamefont
  {H.~Q.}\ \bibnamefont {Pham}}, \bibinfo {author} {\bibfnamefont
  {A.}~\bibnamefont {Pulkin}}, \bibinfo {author} {\bibfnamefont
  {W.}~\bibnamefont {Purwanto}}, \bibinfo {author} {\bibfnamefont {P.~J.}\
  \bibnamefont {Robinson}}, \bibinfo {author} {\bibfnamefont {E.}~\bibnamefont
  {Ronca}}, \bibinfo {author} {\bibfnamefont {E.~R.}\ \bibnamefont
  {Sayfutyarova}}, \bibinfo {author} {\bibfnamefont {M.}~\bibnamefont
  {Scheurer}}, \bibinfo {author} {\bibfnamefont {H.~F.}\ \bibnamefont
  {Schurkus}}, \bibinfo {author} {\bibfnamefont {J.~E.~T.}\ \bibnamefont
  {Smith}}, \bibinfo {author} {\bibfnamefont {C.}~\bibnamefont {Sun}}, \bibinfo
  {author} {\bibfnamefont {S.-N.}\ \bibnamefont {Sun}}, \bibinfo {author}
  {\bibfnamefont {S.}~\bibnamefont {Upadhyay}}, \bibinfo {author}
  {\bibfnamefont {L.~K.}\ \bibnamefont {Wagner}}, \bibinfo {author}
  {\bibfnamefont {X.}~\bibnamefont {Wang}}, \bibinfo {author} {\bibfnamefont
  {A.}~\bibnamefont {White}}, \bibinfo {author} {\bibfnamefont {J.~D.}\
  \bibnamefont {Whitfield}}, \bibinfo {author} {\bibfnamefont {M.~J.}\
  \bibnamefont {Williamson}}, \bibinfo {author} {\bibfnamefont
  {S.}~\bibnamefont {Wouters}}, \bibinfo {author} {\bibfnamefont
  {J.}~\bibnamefont {Yang}}, \bibinfo {author} {\bibfnamefont {J.~M.}\
  \bibnamefont {Yu}}, \bibinfo {author} {\bibfnamefont {T.}~\bibnamefont
  {Zhu}}, \bibinfo {author} {\bibfnamefont {T.~C.}\ \bibnamefont {Berkelbach}},
  \bibinfo {author} {\bibfnamefont {S.}~\bibnamefont {Sharma}}, \bibinfo
  {author} {\bibfnamefont {A.~Y.}\ \bibnamefont {Sokolov}},\ and\ \bibinfo
  {author} {\bibfnamefont {G.~K.-L.}\ \bibnamefont {Chan}},\ }\href
  {https://doi.org/10.1063/5.0006074} {\bibfield  {journal} {\bibinfo
  {journal} {The Journal of Chemical Physics}\ }\textbf {\bibinfo {volume}
  {153}},\ \bibinfo {pages} {024109} (\bibinfo {year} {2020})}\BibitemShut
  {NoStop}%
\bibitem [{\citenamefont {Iskakov}\ \emph {et~al.}(2020)\citenamefont
  {Iskakov}, \citenamefont {Yeh}, \citenamefont {Gull},\ and\ \citenamefont
  {Zgid}}]{Iskakov20}%
  \BibitemOpen
  \bibfield  {author} {\bibinfo {author} {\bibfnamefont {S.}~\bibnamefont
  {Iskakov}}, \bibinfo {author} {\bibfnamefont {C.-N.}\ \bibnamefont {Yeh}},
  \bibinfo {author} {\bibfnamefont {E.}~\bibnamefont {Gull}},\ and\ \bibinfo
  {author} {\bibfnamefont {D.}~\bibnamefont {Zgid}},\ }\href
  {https://doi.org/10.1103/PhysRevB.102.085105} {\bibfield  {journal} {\bibinfo
   {journal} {Phys. Rev. B}\ }\textbf {\bibinfo {volume} {102}},\ \bibinfo
  {pages} {085105} (\bibinfo {year} {2020})}\BibitemShut {NoStop}%
\bibitem [{\citenamefont {Yeh}\ \emph {et~al.}()\citenamefont {Yeh},
  \citenamefont {Chen}, \citenamefont {Fei}, \citenamefont {Gull},\ and\
  \citenamefont {Zgid}}]{Yeh21}%
  \BibitemOpen
  \bibfield  {author} {\bibinfo {author} {\bibfnamefont {C.}~\bibnamefont
  {Yeh}}, \bibinfo {author} {\bibfnamefont {Y.}~\bibnamefont {Chen}}, \bibinfo
  {author} {\bibfnamefont {J.}~\bibnamefont {Fei}}, \bibinfo {author}
  {\bibfnamefont {E.}~\bibnamefont {Gull}},\ and\ \bibinfo {author}
  {\bibfnamefont {D.}~\bibnamefont {Zgid}},\ }\href@noop {} {\bibinfo {title}
  {Wannier interpolation of the self-energy in real materials systems}},\
  \bibinfo {howpublished} {in preparation}\BibitemShut {NoStop}%
\bibitem [{\citenamefont {Werner}\ \emph {et~al.}(2009)\citenamefont {Werner},
  \citenamefont {Gull},\ and\ \citenamefont {Millis}}]{Werner09}%
  \BibitemOpen
  \bibfield  {author} {\bibinfo {author} {\bibfnamefont {P.}~\bibnamefont
  {Werner}}, \bibinfo {author} {\bibfnamefont {E.}~\bibnamefont {Gull}},\ and\
  \bibinfo {author} {\bibfnamefont {A.~J.}\ \bibnamefont {Millis}},\ }\href
  {https://doi.org/10.1103/PhysRevB.79.115119} {\bibfield  {journal} {\bibinfo
  {journal} {Phys. Rev. B}\ }\textbf {\bibinfo {volume} {79}},\ \bibinfo
  {pages} {115119} (\bibinfo {year} {2009})}\BibitemShut {NoStop}%
\bibitem [{\citenamefont {Georges}\ \emph {et~al.}(2013)\citenamefont
  {Georges}, \citenamefont {Medici},\ and\ \citenamefont
  {Mravlje}}]{Georges13}%
  \BibitemOpen
  \bibfield  {author} {\bibinfo {author} {\bibfnamefont {A.}~\bibnamefont
  {Georges}}, \bibinfo {author} {\bibfnamefont {L.~d.}\ \bibnamefont
  {Medici}},\ and\ \bibinfo {author} {\bibfnamefont {J.}~\bibnamefont
  {Mravlje}},\ }\href
  {https://doi.org/10.1146/annurev-conmatphys-020911-125045} {\bibfield
  {journal} {\bibinfo  {journal} {Annual Review of Condensed Matter Physics}\
  }\textbf {\bibinfo {volume} {4}},\ \bibinfo {pages} {137} (\bibinfo {year}
  {2013})}\BibitemShut {NoStop}%
\bibitem [{\citenamefont {Yin}\ \emph {et~al.}(2011)\citenamefont {Yin},
  \citenamefont {Haule},\ and\ \citenamefont {Kotliar}}]{Yin11}%
  \BibitemOpen
  \bibfield  {author} {\bibinfo {author} {\bibfnamefont {Z.~P.}\ \bibnamefont
  {Yin}}, \bibinfo {author} {\bibfnamefont {K.}~\bibnamefont {Haule}},\ and\
  \bibinfo {author} {\bibfnamefont {G.}~\bibnamefont {Kotliar}},\ }\href
  {https://doi.org/10.1038/nmat3120} {\bibfield  {journal} {\bibinfo  {journal}
  {Nature Materials}\ }\textbf {\bibinfo {volume} {10}},\ \bibinfo {pages}
  {932} (\bibinfo {year} {2011})}\BibitemShut {NoStop}%
\end{thebibliography}%

\end{document}